\documentstyle[aps,prd,epsfig]{revtex}
\draft

\begin{document}
\title {The two-pion spectra for the reaction $\pi^-p \to \pi^0\pi^0 n$
at 38 GeV/c pion momentum and combined analysis of the GAMS, Crystal
Barrel and BNL data}
\author{
V. V. Anisovich\thanks{Electronic address: anisovic@thd.pnpi.spb.ru},
A. A. Kondashov\thanks{Electronic address: kondashov@mx.ihep.su},
\frame{Yu. D. Prokoshkin},
S. A.  Sadovsky\thanks{Electronic address: sadovsky@mx.ihep.su},
A. V. Sarantsev\thanks{Electronic address: vsv@hep486.pnpi.spb.ru}}
\address{St.Petersburg Nuclear Physics Institute, Gatchina, 188350,
Russia}
\address{Institute of High Energy Physics, Protvino, Moscow
district, 142284, Russia}
\date{\today}
\maketitle
\newcommand{\beq}{\begin{equation}}
\newcommand{\eeq}{\end{equation}}
\newcommand{\be}{\begin{eqnarray}}
\newcommand{\ee}{\end{eqnarray}}
\newcommand{\inli}{\int\limits}
\newcommand{\qq}{$q\bar q$~}
\newcommand{\fres}{$f_0(980)$~}
\newcommand{\ares}{$a_0(980)$~}

\begin{abstract}

We perform the K-matrix analysis of meson partial waves with
$IJ^{PC} =00^{++}$, $10^{++}$, $02^{++}$, $ 12^{++}$ basing on GAMS
data on  $\pi^-p\to \pi^0\pi^0 n$, $ \eta\eta n$,  $\eta\eta' n$
together with BNL data on $\pi^-p\to K\bar K n$ and  Crystal
Barrel data on  $p\bar p (at \; rest)\to \pi^0\pi^0\pi^0$,
$\pi^0\eta\eta$,  $\pi^0\pi^0\eta$. The positions of the
amplitude poles (physical resonances) are determined as well as
the positions of the K-matrix poles (bare states) and the values of
bare state couplings
to two-meson channels. Nonet classification of the determined bare
states is discussed.
\end{abstract}


\narrowtext
\twocolumn
\section{Introduction}

With this paper we complete the K-matrix analysis of  GAMS data
on the reactions $\pi^-p\to \pi^0\pi^0 n$\cite{gams1}, $ \eta\eta n$
\cite{gams2}, $\eta\eta' n$ \cite{gams3} which was started by the
papers \cite{km1100,km1500,km1900}. The K-matrix analysis
gives a rich information about meson states, thus
helping the $q\bar q$ state  classification  and the search for
exotic mesons.  However for the restoration of the K-matrix amplitude,
one needs to study a full set of  open channels with sufficiently
high statistics.  It is the  reason to include in our fit the
data on $\pi^-p\to K\bar K n$ \cite{bnl} and  $p\bar p (at \; rest)\to
\pi^0\pi^0\pi^0$, $\pi^0\eta\eta$,  $\pi^0\pi^0\eta$ \cite{cbc}.

K-matrix poles which are a subject of the present
consideration differ from the amplitude poles in two points:

(i) The states corresponding to the K-matrix poles do not contain
 components with real mesons which are inherent in resonances. The
absence of a cloud of  real mesons allows one to refer conventionally
to these states  as the bare ones \cite{km1500,km1900}.

(ii) Due to the transition $bare\;  state \to real \; mesons\to bare \;
state$, the observed resonances are mixtures of the bare states. So,
for the quark systematics, the bare states are primary objects
rather than  resonances.

Coupling constants $bare\;  state \to real \; mesons$ are responsible
not only for the mixing of states but for the resonance decays as well;
the relations between couplings  allow one to
restore the quark/gluon content of  bare states
\cite{glp,aas-z}.

The paper is organized as follows.

In Sec. II we introduce a set of
formulae which are used in the data fit. We present the S- and
D-wave K-matrix amplitudes for the mass-on-shell reactions $\pi\pi \to
\pi\pi , K\bar K , \eta\eta , \eta\eta'$  together with  those for the
mass-off-shell  pion in the initial state: $\pi\pi (t)$ with $t\ne
m_{\pi}^2$.  The K-matrix formulae for the final state interaction in
the three-meson production process, $p\bar p \to three \; mesons$, are
presented as well.

In Sec. III we write down the couplings for the transition $bare \;
state\to \;two \; pseudoscalars$, with the imposed quark-combinatorics
constraints both for $q\bar q$-states (isoscalar and isovector)
and for the glueball. Restoration  of couplings in  the
 fit allows us to determine the quark content of isoscalar states
and to find a candidate for the glueball.

Mesons which belong to the same $q\bar q$ nonet have
approximately equal masses; they also have approximately equal
decay couplings.  Besides, flavour wave functions for isoscalars
of the same nonet are orthogonal. In Sec. IV we present
the results of the fit with the imposed nonet-classification
constraints.  Fit of the $00^{++}$ wave confirms  the  result
of Ref. \cite{km1900}, while for the $02^{++}$, $10^{++}$, $12^{++}$
waves the K-matrix representation of amplitudes in the mass region
below 1900 MeV is done for the first time. The restored bare states,
together with those found in the $K\pi$ S-wave K-matrix analysis
\cite{alan}, allow us to construct the $1^3P_0q\bar q$ and
$1^3P_2q\bar q$ nonets unambigously, and for the $2^3P_0q\bar q$ nonet
two variants are possible which differ in the mass of the lightest
scalar/isoscalar state.

The origin of the lightest scalars, $f_0(980)$ and $a_0(980)$,
is crucial for the nonet classification. These states are
located near the $K\bar K$ threshold and give rise to the question
whether these states are hadronic  $K\bar K$ molecules. In Sec. 5 we
present arguments based on the direct GAMS measurements
together with the results of the performed K-matrix fit that the bare
states from which $f_0(980)$ and $a_0(980)$ originate have $q\bar q$
nature.

Short summary is given in Sec. 6.

\section{Experimental data and K-matrix amplitude}

Here we briefly introduce  the fitted data and set out the
K-matrix formulae used for the data analysis.

\subsection{Experimental data}

Simultaneous analysis of meson spectra in the channels
$IJ^{PC}=00^{++}$, $10^{++}$, $02^{++}$ and $12^{++}$
is performed on the basis of the following data set:\\
(1) GAMS data on the S-wave two-meson  production in the reactions
 $\pi p\to \pi^0\pi^0 n$,
$\eta\eta n$ and $\eta\eta' n$
at small nucleon momenta transferred, $|t|<0.2$ (GeV/c)$^2$
\cite{gams1,gams2,gams3};\\
(2) GAMS data on the $\pi\pi$ S-wave production in the reaction
 $\pi p\to \pi^0\pi^0 n$
at large momenta transferred,  $0.30<|t|<1.0$ (GeV/c)$^2$
\cite{gams1};\\
(3) GAMS data on the $\pi\pi$ D-wave production in the reaction
$\pi p\to \pi^0\pi^0 n$,
 at small and large $|t|$, $0<|t|<0.5$ (GeV/c)$^2$
\cite{gams3}; \\
(4) BNL data on $\pi N\to K\bar K N$ \cite{bnl};\\
(5)  Crystal Barrel data on
$p\bar p~(at~rest)\to \pi^0\pi^0\pi^0$, $\pi^0\pi^0\eta$,
$\pi^0\eta\eta$ \cite{cbc};\\

\subsection{K-matrix amplitude and analyticity}

The K-matrix technique is used for the description
of the two-meson coupled channels:
\beq
A= K(I-i\hat{\rho} K)^{-1},
\label{1}
\eeq
where $K$ is $n\times n$ matrix, where $n$ is the number of channels
under consideration
and $I$ is unit matrix. The phase
space matrix is diagonal:  $\hat{\rho}_{ab}=\delta_{ab}\rho_a$. The
phase
space factor $\rho_a$ is responsible for the threshold singularities
of the amplitude: to keep the amplitude analytic in the
physical region under consideration we use analytic continuation for
$\rho_a$ below threshold.  For example, the $\eta\eta$ phase space
factor $\rho_{a}=(1-4m_\eta^2/s)^{1/2}$ is equal to
$i(4m_\eta^2/s-1)^{1/2}$  below the $\eta\eta$ threshold
($s$ is two-meson invariant energy squared). The phase
space factors we use lead to false kinematic singularities at $s=0$
(in all factors)
and at $s = (m_{\eta'}-m_\eta)^2$ (in the $\eta\eta'$ space factor),
but these false singularities which are standard for the K-matrix
approach are rather distant from the investigated physical region.

For the multimeson phase volume in the isoscalar sector, we
use the four-pion phase space defined as either $\rho \rho$
or  $\sigma \sigma$ phase space, where $\sigma$ denotes the S-wave
$\pi\pi$ amplitude below  1.2 GeV.  The result does not
practicaly depend on whether we use $\rho \rho$ or $\sigma \sigma$
state for the description of multimeson channel: below we provide the
formulae and the values of the obtained parameters for the $\rho \rho$
case, for which the fitted expressions  are less cumbersome.
The multimeson phase space in the sector $I=1$ is taken in the form
which, in its low-energy part, simulates the $a_0\rho$ phase space.

\subsection{Isoscalar/scalar, $00^{++}$, partial wave}

For the S-wave interaction in the isoscalar sector,
we use the parametrization similar
to that of Ref.  \cite{km1900}:
$$K_{ab}^{00}(s)=$$
\be
\left ( \sum_\alpha \frac{g^{(\alpha)}_a
g^{(\alpha)}_b}
{M^2_\alpha-s}+f_{ab}
\frac{1\;\mbox{GeV}^2+s_0}{s+s_0} \right )\;
\frac{s-m_\pi^2/2}{s}\;\;,
\label{2}
\ee
where $K_{ab}^{IJ}$ is a 5$\times $5 matrix ($a,b$ = 1,2,3,4,5), with
the following notations for meson states: 1 = $\pi\pi$, 2 = $K\bar K$,
3 = $\eta\eta$, 4 = $\eta\eta'$ and 5 = {\it multimeson states}
(four-pion state mainly at $\sqrt{s}<1.6\; \mbox{GeV}$).
The $g^{(\alpha)}_a$ is a coupling constant of the bare state
$\alpha$ to the meson channel; the parameters $f_{ab}$ and $s_0$
describe the smooth part of the K-matrix elements ($s_0>1.5$ GeV$^2$).
 We use  the factor $(s-m_\pi^2/2)/s$ to suppress the influence of the
false kinematic singularity at $s=0$ in the amplitude near the $\pi\pi$
threshold.

The  phase space matrix elements  are equal to:
\beq
\rho_a(s)=\sqrt{\frac{s-4m_a^2}{s}}\qquad , \qquad a=1,2,3,
\label{3}
\eeq
where $m_1=m_\pi$, $m_2=m_K$, $m_3=m_\eta$, and
\be
\rho_4(s)=\left \{
\begin{array}{cl}
\rho_{41}\;\;\mbox{at} \;\; s>(m_\eta-m_{\eta'})^2\\
\rho_{42}\;\;\mbox{at} \;\; s<(m_\eta-m_{\eta'})^2
\end{array} \right. , \nonumber
%\label{4}
\ee
\be
\rho_{41}=&&\sqrt{\left (1-\frac{(m_\eta+m_{\eta'})^2}{s}\right )
\left (1-\frac{(m_\eta-m_{\eta'})^2}{s}\right )} \;,\nonumber \\
\rho_{42}=&&0.
\ee
The multimeson phase space factor is defined as
\be
\rho_5(s)=\left \{ \begin{array}{cl}
\rho_{51}\;\;\mbox{at} \;\;s<1\;\mbox{GeV}^2\\
\rho_{52}\;\;\mbox{at} \;\;s>1\;\mbox{GeV}^2
 \end{array} \right . , \nonumber
%\label{5}
\ee
\be
\rho_{51}=&&\rho_0\int\frac{ds_{1}}{\pi}\int\frac{ds_{2}}{\pi}
\nonumber \\
&&\times M^2\Gamma(s_{1})\Gamma(s_{2})
\sqrt{(s+s_{1}-s_{2})^2-4ss_{1}} \nonumber \\
&&\times s^{-1}\left [(M^2-s_{1})^2+M^2\Gamma^2(s_{1})\right
]^{-1}\nonumber \\
&&\times \left
[(M^2-s_{2})^2+M^2\Gamma^2(s_{2})\right ]^{-1} ,\nonumber \\
\rho_{52}=&&1\,.
\ee
Here $s_1$ and $s_2$ are the two-pion energies
squared, $M$ is $\rho$-meson mass and
$\Gamma(s)$ is its energy-dependent width,
$\Gamma(s)=\gamma\rho_1^3(s)$. The factor $\rho_0$ provides the
continuity of $\rho_5(s)$  at $s=1$ GeV$^2$.

The following formulae describe  $\pi\pi$, $\eta\eta$ and
$\eta\eta'$ production amplitudes due to pion $t$-channel exchange:
\be
A_{\pi N\to N b}=&&
N(\bar\Psi_N\gamma_5\Psi_N)F_N(t)D(t)\tilde K_{\pi\pi(t),a}  \nonumber \\
&&\times(1-i\rho  K)^{-1}_{ab}\, , \qquad
b=\pi\pi,\eta\eta,\eta\eta'\;\;,  \nonumber \\
\tilde K_{\pi\pi(t),a}=&&\left (
\sum_\alpha \frac{\tilde g^{(\alpha)}(t)
g^{(\alpha)}_a} {M^2_\alpha-s}+\tilde f_{a}(t)\;
\frac{1\;\mbox{GeV}^2+s_0}{s+s_0} \right )  \nonumber \\
 &&\times \left (s-m_\pi^2/2\right )/s\;.
\label{6}
\ee
Here $N$ is a normalization factor, $F_N(t)$ is the nucleon form
factor, and $D(t)$ is the pion propagator:
\be
F_N(t)=&&\left [
\frac{\tilde\Lambda-m_\pi^2} {\tilde\Lambda-t}\right ]^4\;,
\nonumber \\
D(t)=&&(m_\pi^2-t)^{-1}\;\;,  \nonumber \\
\tilde g^{(\alpha)}(t)=&&
g^{(\alpha)}_1+(1-\frac{t}{m_\pi^2})\,
(\Lambda_g-\frac{t}{m_\pi^2})g'^{(\alpha)}\;\;,  \nonumber \\
\tilde f_{a}(t)=&&f_{1a}+(1-\frac{t}{m_\pi^2})
(\Lambda_f-\frac{t}{m_\pi^2})f'_{a}\;\;,
\label{7}
\ee
where $\Lambda$'s, $g'$ and $f'$ are the fitted parameters.

\subsection{Isoscalar/tensor, $02^{++}$, partial wave}

The D-wave interaction in the isoscalar sector
is parametrized by the 4$\times $4 K-matrix where
1 = $\pi\pi$, 2 = $K\bar K$,
3 = $\eta\eta$ and 4 = {\it multimeson states}:
\be
K_{ab}^{0,2}(s)=&&D_a(s)\left ( \sum_\alpha \frac{g^{(\alpha)}_a
g^{(\alpha)}_b}{M^2_\alpha-s} \right . \nonumber \\
&&\left .+f_{ab}\frac{1\,\mbox{GeV}^2+s_2}{s+s_2} \right )
D_b(s)\;.
\label{8}
\ee
Factor $D_a(s)$ stands for the D-wave centrifugal barrier.
We take this factor in the following form:
\be
D_a(s)=\frac{k_a^2}{k_a^2+3/r_a^2},\;\;a=1,2,
\ee
where $k_a=\sqrt{s/4-m_a^2} $ is the momentum of the decaying
meson in the c.m. frame  of the resonance. For the multimeson decay,
the factor $D_5(s)$ is taken to be equal to 1. The phase space factors
used are the same as those for the isoscalar S-wave channel.

\subsection{Isovector/scalar, $10^{++}$, and isovector/tensor,
$12^{++}$, partial waves}

For the amplitude in the isovector/scalar and isovector/tensor
channels, we use the 4$\times $4 K-matrix with
1 = $\pi\eta$, 2 = $K\bar K$, 3 = $\pi\eta'$ and
4 = {\it multimeson states}:
\be
K_{ab}^{1J}(s)=&&D_a(s)
 \left ( \sum_\alpha \frac{g^{(\alpha)}_a
g^{(\alpha)}_b}{M^2_\alpha-s} \right . \nonumber \\
&&\left . +f_{ab}\frac{1.5\; \mbox{GeV}^2+s_1}{s+s_1} \right )D_b(s)\;.
\label{10}
\ee
Here $J=0,2$; the factors $D_a(s)$ are equal to 1 for the $10^{++}$
amplitude, while for the
D-wave partial amplitude the factor $D_a(s)$ is taken in the form:
\be
D_a(s)=&&\frac{k_a^2}{k_a^2+3/r_3^2}, \;\; a=1,2,3,  \nonumber \\
D_4(s)=&&1\; .
\label{11}
\ee
The elements of the phase space matrix in the isovector sector
are defined as
\be
\rho_1(s)=&&\left \{
\begin{array}{cl}
\rho_{11} \qquad \mbox{at} \qquad s>(m_\eta-m_{\pi})^2  \\
\rho_{12} \qquad \mbox{at} \qquad s<(m_\eta-m_{\pi})^2  \\
\end{array} \right .,\nonumber
\ee
\be
\rho_{11}=&&\sqrt{\left (1-\frac{(m_\eta+m_{\pi})^2}{s}\right )
\left (1-\frac{(m_\eta-m_{\pi})^2}{s}\right )}\,, \nonumber \\
\rho_{12}=&&0\,,
\label{12}
\ee
\be
\rho_2(s)=&&\sqrt{\frac{s-4m_K^2}{s}}\, ,
\label{13}
\ee
\be
\rho_3(s)=&&\left \{
\begin{array}{cl}
\rho_{31} \qquad \mbox{at} \qquad s>(m_\eta'-m_{\pi})^2\\
\rho_{32} \qquad \mbox{at} \qquad s<(m_\eta'-m_{\pi})^2\\
\end{array}  \right . ,
\nonumber
\ee
\be
\rho_{31}=&&\sqrt{\left (1-\frac{(m_\eta'+m_{\pi})^2}{s}\right )
\left (1-\frac{(m_\eta'-m_{\pi})^2}{s}\right )}\,, \nonumber \\
\rho_{32}=&&0\,.
\label{14}
\ee
The multimeson phase space factor $\rho_4 (s)$ is taken in the form
which simulates
the $\rho a_0$ phase space factor
below $s=2.25$ GeV$^2$:
\be
\rho_4(s)=\left \{
\begin{array}{cl}
\rho_{41}&  \qquad \mbox{at} \;\;\;\;  (m_\eta+3m_\pi)^2<s<2.25
\,\mbox{GeV}^2\,\\
\rho_{42}&  \qquad \mbox{at} \qquad  \; \qquad   s>2.25 \,\mbox
{GeV}^2\,\\ \rho_{43}&  \qquad \mbox{at}  \qquad \; \qquad
s<(m_\eta+3m_{\pi})^2  \end{array} \right . ,
\nonumber
%\label{15}
\ee
\be
\rho_{41}=&&\left (\frac{1-(m_\eta+3m_\pi)^2/s}
{1-(m_\eta+3m_\pi)^2/2.25\, \mbox{GeV}^2} \right )^{5/2} \,,
\nonumber \\
\rho_{42}=&&1 \,,  \nonumber \\
\rho_{43}=&&0 \; .
\ee
\subsection{Three-meson production amplitudes}

The amplitudes $p\bar p~(at~rest)\to
\pi^0\pi^0\pi^0$,
$\pi^0\eta\eta$ which correspond to the
production of the two-meson
isoscalar states are equal to:
$$
A_{p\bar p\to {\rm three\; mesons}}=A_1(23)+A_2(13)+A_3(12) \; ,
\label{15a}
$$
where the amplitude $A_k(s_{ij})$ stands for  diagrams with an
interaction of particles in the intermediate states and  the last
interaction being of the particles $i$ and $j$, while the particle
$k$ is a spectator.  We suppose, as in the previous papers
\cite{km1500,km1900}, that $p\bar p$ annihilates at rest from the
$^1S_0$-level.  The following form is used for the two-particle
interaction block:
\be
A_1(23)=&&\sum\limits_{J=0,2} X_J(23)\;\tilde
K_{p\bar p\pi,a}^{0J} (s_{23}) \nonumber \\
&&\times \left(1-i \hat\rho
K^{0J}(s_{23})\right)^{-1}_{ab}\; .
\label{16}
\ee
Here
$b=\pi^0\pi^0$ stands for $\pi^0\pi^0\pi^0$ production, and $b=\eta\eta$
for $\pi^0\eta\eta$. The centrifugal barrier factor $X_J$ is equal to 1
for the production of the S-wave resonance. For the D-wave resonance
production, this factor is:
\be
X_2(23)=\frac
12(3\cos^2\Theta_{12}-1) \frac{p_1^2}{p_1^2+3/R^2},
\label{17}
\ee
where $\Theta_{12}$ is the angle between particles 1 and 2 in the
rest frame of the particles 2 and 3,  $p_1$ is the momentum of the
particle 1 in this frame
and $R$ characterizes the annihilation radius.
The $\tilde K$-matrices which describe the prompt resonance
production
 in the $p\bar p$ annihilation have  the following
form:
\be
\tilde K_{p\bar p\pi,a}^{00}(s_{ij})=&&\left (
\sum_\alpha \frac{\Lambda^{(\alpha)}_{p\bar p\pi}[00]
g^{(\alpha)}_a}
{M^2_\alpha-s_{ij}} +\phi_{p\bar p\pi,a}[00]\; \right.
\nonumber \\
&&\left . \times
\frac{1\; \mbox{GeV}^2+s_0}{s_{ij}+s_0} \right)
\left (\frac{s_{ij}-m_\pi^2/2}{s_{ij}}\right )\;,
\label{18}
\ee
\be
\tilde K_{p\bar p\pi,a}^{02}(s_{ij})=&&\left (
\sum_\alpha \frac{\Lambda^{(\alpha)}_{p\bar p\pi}[02] g^{(\alpha)}_a}
{M^2_\alpha-s_{ij}}
+\phi_{p\bar p\pi,a}[0,2]\;\right .
\nonumber \\
&& \left. \times
\frac{1\; \mbox{GeV}^2+s_0}{s_{ij}+s_0} \right)
D_a(s_{ij})\,.
\label{19}
\ee
The $\pi\pi\pi$ production amplitude is completely described by eqs.
(\ref{16})-(\ref{19}) because of the amplitude symmetry under
the rotation of pion indices $i,j,k$.

The part of the amplitude $p\bar p~(at~rest)\to \pi \pi^0\pi^0$, which
corresponds to the production of  isoscalar
resonances, reads:
\be
A_1(23)=&&\sum\limits_{J=0,2}X_J(23)\;\tilde K_{p\bar p\pi,a}^{0J}
(s_{23})
\nonumber \\
&&\left(1-i \rho K^{0J}(s_{23})\right)^{-1}_{ab}
\;\;,\quad b=\pi^0 \pi^0 \;\;,
\nonumber \\
\ee
where
\be
\tilde K_{p\bar p\eta,a}^{00}(s_{ij})=&&\left (
\sum_\alpha \frac{\Lambda^{(\alpha)}_{p\bar p\eta}[00] g^{(\alpha)}_a}
{M^2_\alpha-s_{ij}}+\phi_{p\bar p\eta,a}[00]\; \right .
\nonumber \\
&& \left . \times \frac{1\; \mbox{GeV}^2+s_0}{s_{ij}+s_0} \right)\;
\left (\frac{s_{ij}-m_\pi^2/2}{s_{ij}}\right )\;.
\label{20}
\ee
Parameters $\Lambda^\alpha_{p\bar p\pi}[0J]$ and
$\phi_{p\bar p\pi}[0J]$
(or $\Lambda^\alpha_{p\bar p\eta}[0J]$ and $\phi_{p\bar p\eta}[0J]$)
may be complex magnitudes with different phases due to the three
particle interactions.

The part of the amplitude,
which corresponds to the production of the
isovector resonances in the
reaction $p\bar p~(at~rest)\to \eta\eta\pi^0$,
 is written as $ A_1(23)+A_2(13)$ and
\be
A_2(13)=&&\sum\limits_{J=0,2}
X_J(13)\;\tilde K_{p\bar p\eta,a}^{1J}
(s_{13})
\nonumber \\
&&\times \left( 1-i \hat\rho
K^{1J}(s_{13})\right)^{-1}_{ab}, \qquad b=\eta\pi^0 \; ,
\label{21}
\ee
where
\be
\tilde K_{p\bar p\eta,a}^{1J}(s_{ij})=&&\left (
\sum_\alpha \frac{\Lambda^{(\alpha)}_{p\bar p\eta}[1J]
g^{(\alpha)}_a}
{M^2_\alpha-s_{ij}} \right.
\nonumber \\
&&\left .+\phi_{p\bar p\eta,a}[1J]\;
\frac{1\; \mbox{GeV}^2+s_1}{s_{ij}+s_1} \right)D_a(s_{ij})\;.
\label{22}
\ee
The production of isovector resonances in the reaction
$p\bar p~(at~rest)\to \pi^0\pi^0\eta$ has the  form
$A_1(23)+A_2(13)$ and
\be
A_2(13)=&&\sum\limits_{J=0,1,2}
X_J(13)\;\tilde K_{p\bar p\pi,a}^{1J}
(s_{13})
\nonumber \\
&&\times \left(1-i \hat\rho K^{1J}(s_{13}\right)^{-1}_{ab},
\qquad b=\eta\pi^0 \; ,
\label{23}
\ee
where $\tilde K_{p\bar p\pi,a}^{1J}$
is given by  Eq. (\ref{22}) with the replacement
$\Lambda^{(\alpha)}_{p\bar p\eta}[1J] \to
\Lambda^{(\alpha)}_{p\bar p\pi}[1J] $   and
$\phi_{p\bar p\eta,a}[1J] \to \phi_{p\bar p\pi,a}[1J]  $.

\section{Quark-combinatoric rules for the decay couplings
and the \qq content of mesons}

\begin{figure}
\epsfig{file=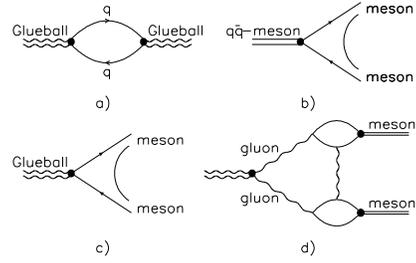,width=7cm}
\caption{Quark-antiquark loop diagram which determines the glueball width
(a); diagrams for the decay of a
$q\bar q$-meson (b) and a glueball (c),(d)
into two $q\bar q$-meson states.}
\label{fig1}
\end{figure}

The decay couplings of the
$q\bar q$-meson and glueball to the two mesons are determined
by the diagrams with $q\bar q$-pairs
produced by  gluons. Figs.~\ref{fig1}(b),(c)
provide an example of diagrams which contribute to the leading terms in
the 1/N expansion \cite{t'h} and Fig.~\ref{fig1}(d) is an example of
diagrams for the next-to-leading contribution. The production of soft
$q\bar q$ pairs by gluons violates flavour symmetry, with the following
ratios of the production probabilities:  \be u\bar u:d\bar d:s\bar
s=1:1:\lambda \; , \label{3.1} \ee and $\lambda=0.4-0.8$ \cite{lambda}.
In our fit we fix $\lambda=0.6$.

We calculate the ratios of the decay coupling constants in the
framework of the quark combinatoric rules which were previously suggested
for the high energy hadron production \cite{as-bf}
and then extended for hadronic $J/\Psi$ decays \cite{wol}.
Calculations
of the decay coupling constants for the glueball and isoscalar/scalar
$q\bar q$-mesons were performed in Refs.  \cite{km1500,glp,aas-z}.  The
decay couplings for isoscalar and isovector mesons are given in Tables
\ref{table1} and \ref{table2}.

Isoscalar meson decay couplings depend on the nonstrange/strange
component ratio of the decaying meson
given by the mixing angle $\Phi$:
\be
\psi^{\mbox{flavour}}_{\mbox{decaying} \; \mbox{meson}}=
n\bar n Cos\Phi+s\bar s Sin\Phi
\; .
\label{3.2}
\ee
Here $n\bar n=(u\bar u+d\bar d)/\sqrt 2 $. It allows us to restore
$\Phi$ and at the same time to determine the decay couplings.

The glueball decay couplings
obey the same ratios as the isoscalar/scalar $q\bar
q$-meson couplings with the  mixing angle equal to:
\be
\Phi = \Phi_{\mbox{glueball}}\; \; \;,
\tan\Phi_{\mbox{glueball}}=\sqrt{\lambda/2}\; .
\label{3.3}
\ee
It follows from the two-stage decay of the glueball \cite{aas-z}, see
Fig.~\ref{fig1}(c): an intermediate $q\bar q$-state in the glueball
decay is a mixture of  $n\bar n$ and $s\bar s$ quarks produced in
proportion given by Eq. (\ref{3.1}).  We fix $\Phi_{glueball}=25^o\pm
5^o$.

The coincidence of the glueball decay couplings with those of $q\bar
q$ meson at $\Phi=\Phi_{glueball}$ points out that there is no simple
signature of a determination of glueball state: searching for the
glueball, we should perform a full $q\bar q$-classification of
mesons, thus an existence of an extra state for the $q\bar
q$-classification is an indication of the exotics.

The normalization in Table \ref{table1} is done
in such a way that for the glueball decay
the summation of couplings squared over all channels is proportional to
the probability to produce a two-quark pair, $(2+\lambda)^2$, see Eq.
(\ref{3.1}). So,
\be \sum\limits_{channels}
G^2(c)I(c)=&&\frac{1}{2}G^2(2+\lambda)^2, \nonumber \\
\sum\limits_{channels}
g_G^2(c)I(c)=&&\frac{1}{2}g_G^2(2+\lambda)^2.
\label{24}
\ee
Here $I(c)$ is the identity factor and $c=\pi^0\pi^0$, $\pi^+\pi^-$,
$K^+K^-$, and so on (see Table \ref{table1}). With this
normalization $ g_G/G\simeq1/N_c$.
The experience of the quark-gluon diagram calculations teaches us that
the factor $1/N_c$ actually leads  to a suppression of
the order of $1/10$: in the fitting procedure we impose a
restriction $| g_G/G|<1/3$.

 The nonet classification
of  isoscalar mesons is based on the following two constraints:
\begin{description}
\item[(1)] The angle difference between
isoscalar nonet partners should be
$90^o$. For this value the corridor $\pm 5^o$ is allowed in
our analysis:
\be
\Phi (1)-\Phi (2)=90^o\pm 5^o \; .
\label{25}
\ee
\item[(2)] Coupling constants $g$ of  Tables \ref{table1} and
\ref{table2} should be approximately equal to each other for all nonet
partners:
\be
g[f_J(1)] \simeq g[f_J(2)]\simeq g[a_J] \simeq g[K_J] \;
.  \label{26}
\ee
\end{description}
 The conventional quark model requires exact coincidence of the
couplings $g$
but the energy dependence of the loop diagram of  Fig.~\ref{fig1}(a),
$B(s)$, may violate this coupling constant equality because of the mass
 splitting inside a nonet. The K-matrix coupling constant  contains an
additional $s$-dependent factor as compared to the coupling of the
N/D-amplitude
\cite{aas-z}: $g^2(K) =  g^2(N/D)/(1+B'(s))$. The factor
$(1+B'(s))^{-1}$ mostly affects the low-$s$ region due to the threshold
and left-hand side singularities of the partial amplitude.
Therefore, the coupling constant equality is mostly violated for the
lightest $00^{++}$ nonet, $1^3P_0$ $q\bar q$. We allow for the members
of this nonet $1\leq g[f_0(1)]/g[f_0(2)]\leq 1.5$.
For the $2^3P_0$ $q\bar q$ nonet members, we put the
two-meson couplings  equal both for isoscalar and isovector
mesons. The equality of coupling constants is also imposed for
tensor resonances.

\section{Description of data and the results}

The performed K-matrix fit gives a good description of the data,
see Figs.~\ref{fig2}-\ref{fig6}.
The $\chi^2$ values for the fit are shown in Table \ref{table3}
and parameters of the fit are presented in Tables
\ref{table4}-\ref{latetable}.  Below we single out the main results of
our fit.

\subsection{$IJ^{PC}=00^{++}$ wave}

The present fit confirms the results obtained in the previous analysis
of the $00^{++}$ wave \cite{km1100,km1500,km1900}.
For the description of the $00^{++}$ wave in the mass region below
2000 MeV, five K-matrix poles are needed (a four-pole amplitude fails
to describe well the data set under consideration). Correspondingly,
five bare states are found:
\be
&&f^{\mbox{bare}}_0(720\pm100), \nonumber \\
\psi^{\mbox{flavour}}=&&(0.45\pm 0.1)n\bar n -(0.89 \pm 0.05)s\bar s \; ,
\nonumber \\
&&f^{\mbox{bare}}_0(1230\pm 50), \nonumber \\
\psi^{\mbox{flavour}}=&&(0.9^{+0.05}_{-0.2})n\bar n
+(0.45^{+0.3}_{-0.1})s\bar s\; ,
\nonumber \\
&&f^{\mbox{bare}}_0(1260\pm 30), \nonumber \\
\psi^{\mbox{flavour}}=&&
(0.93^{+0.02}_{-0.1})n\bar n+(0.37^{+0.2}_{-0.06})s\bar s\; ,
\nonumber \\
&&f^{\mbox{bare}}_0(1600\pm 50), \nonumber \\
\psi^{\mbox{flavour}}=&&
(0.95 \pm 0.05)n\bar n+(0.3^{+014}_{-0.4})s\bar s \; ,
\nonumber \\
&&f^{\mbox{bare}}_0(1810\pm 30), \nonumber \\
\psi^{\mbox{flavour}}=&&\left \{
\begin{array}{cl}
(0.10\pm 0.05)n\bar n+(0.995^{+0.005}_{-0.015})s\bar s, \\
\qquad \mbox{Solution}\; I), \\
(0.67 \pm 0.08)n\bar n-(0.74\pm 0.08)s\bar s, \\
\qquad(\mbox{Solution}\; II).
\end{array} \right .
\label{4.1}
\ee
Experimental data used in the fit do not fix unambigously the flavour
wave function of $f^{\mbox{bare}}_0(1810\pm 30)$:  two solutions
are found for it.

\begin{figure}
\epsfig{file=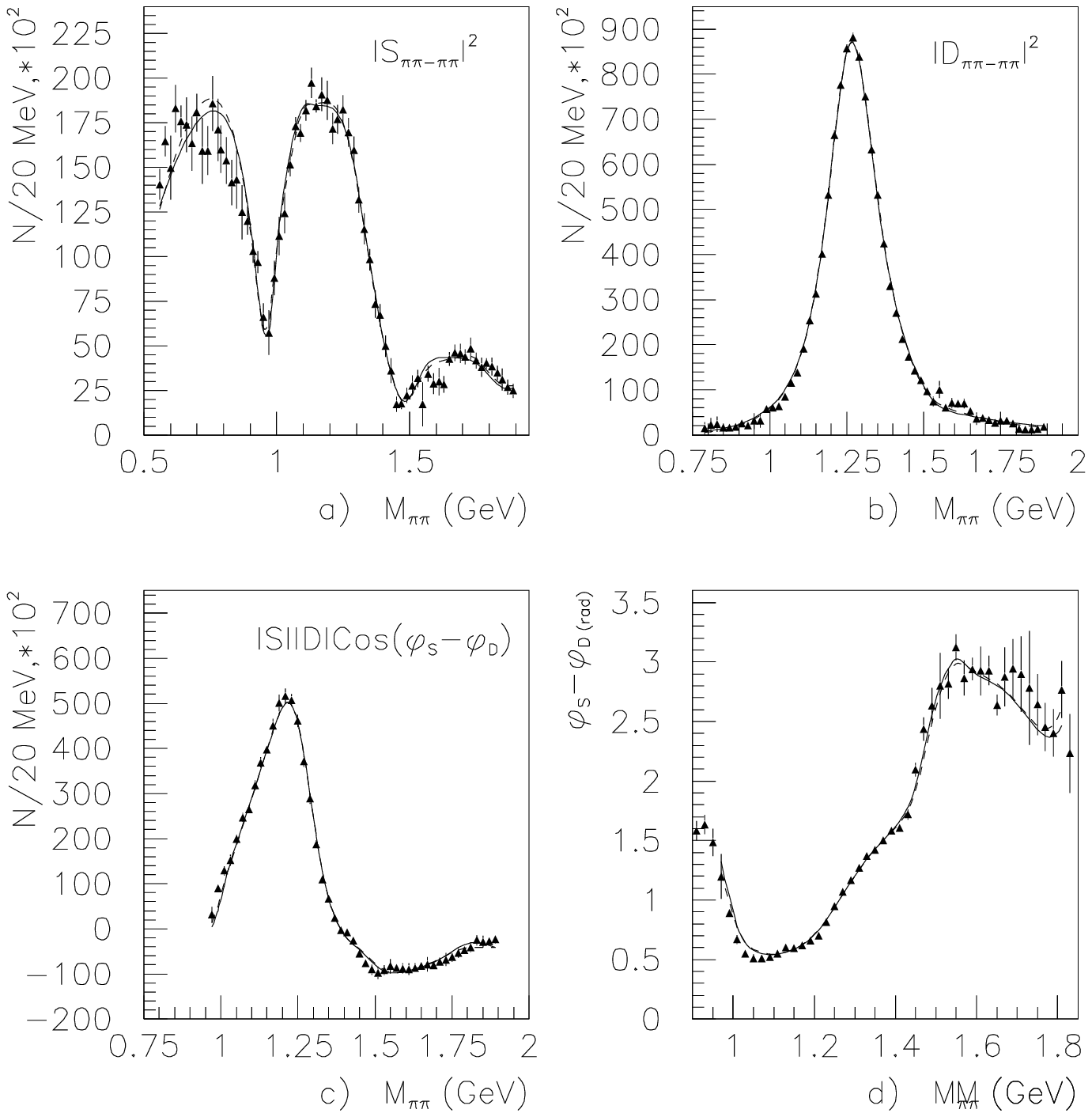,width=16cm}
\caption\protect{The $\pi\pi\to \pi\pi$ $S$-wave amplitude
module squared \cite{gams1} (a),
the $D$-wave amplitude module squared (b),
$SD$-correlator (c) and
the phase difference between S and D-waves (d);
the events are collected at
the momentum transfer
squared $|t|<0.20$ GeV$^2$/c$^2$ \cite{gams1}.
Solid curve corresponds to solution {\bf II-2} and  dashed one
to  solution {\bf I}.}
\label{fig2}
\end{figure}

\begin{figure}
\epsfig{file=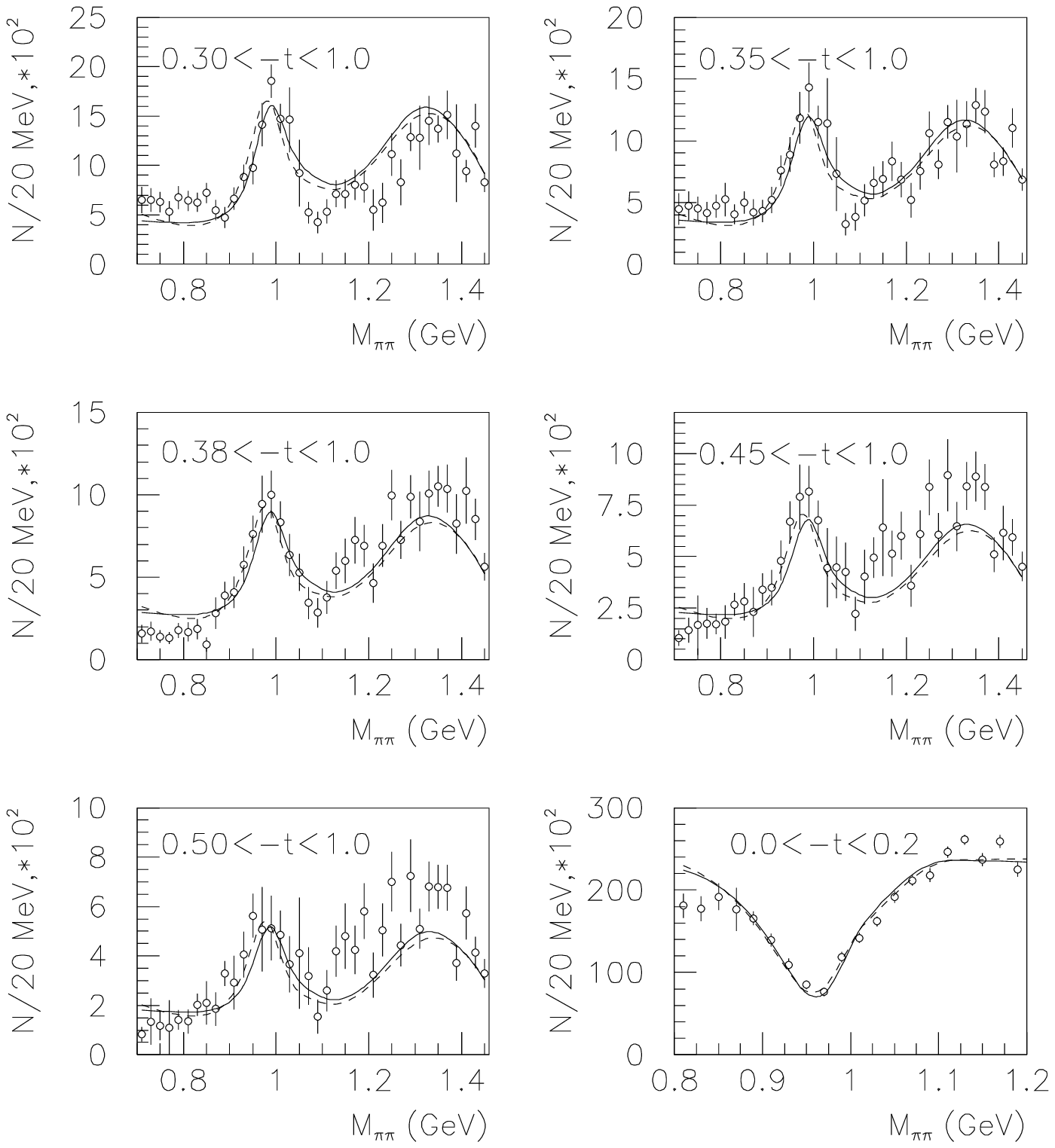,width=16cm}
\caption\protect{ Event
numbers {\it versus}  invariant mass of the $\pi\pi$-system
in the S-wave for different
$t$-intervals in the $\pi^-p\rightarrow \pi^0\pi^0n$ reaction
\cite{gams1}. Solid curves correspond to solution {\bf II-2}
and the dashed curves to solution {\bf I}.}
\label{fig3}
\end{figure}

\begin{figure}
\epsfig{file=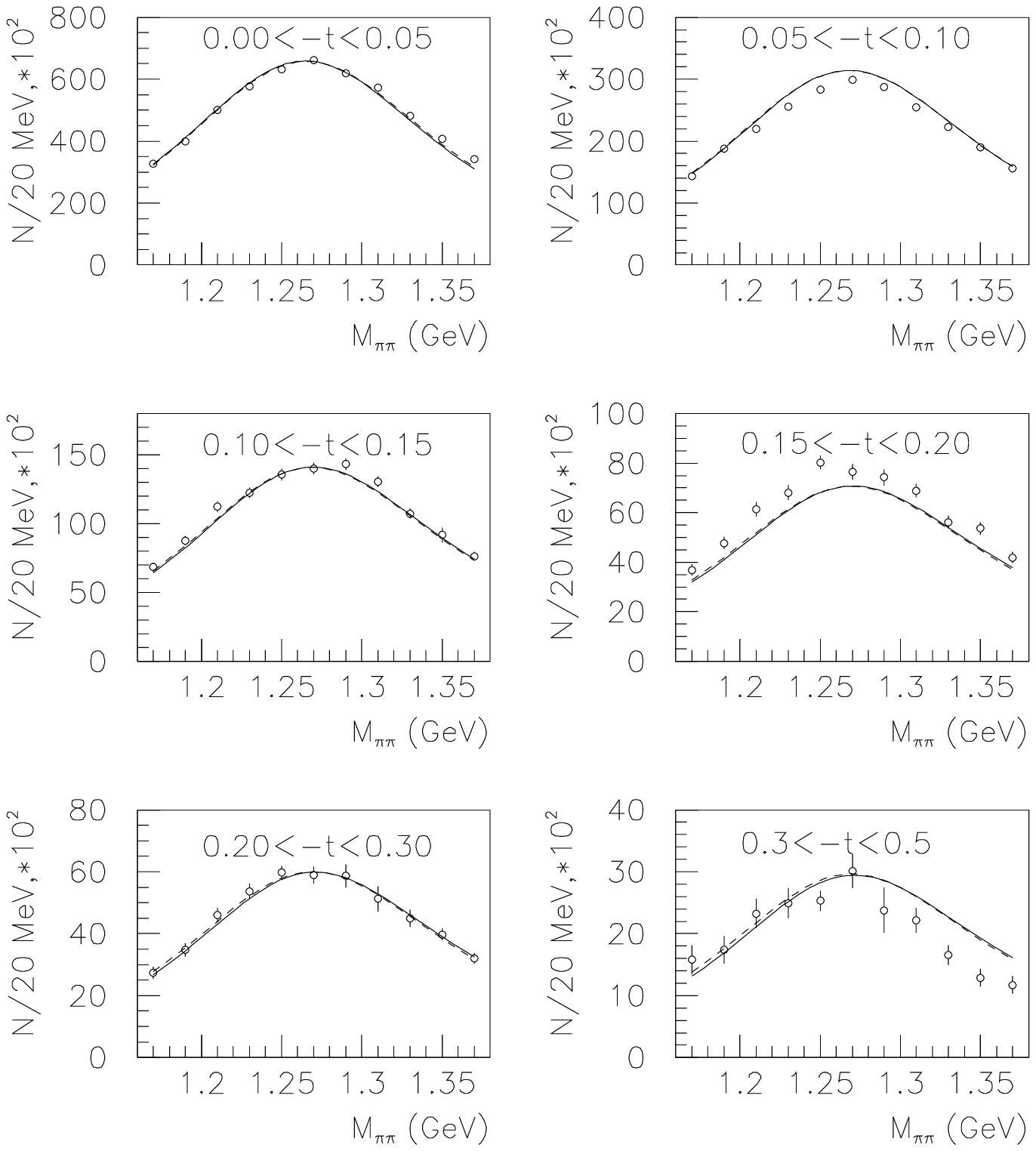,width=16cm}
\caption\protect{Event
numbers {\it versus}  invariant mass of the $\pi\pi$-system
in the D-wave for different
$t$-intervals in the $\pi^-p\rightarrow \pi^0\pi^0n$ reaction
\cite{gams1}. The solid curves correspond to solution {\bf II-2}
and  dashed one to solution {\bf I}.}
\label{fig4}
\end{figure}

\begin{figure}
\epsfig{file=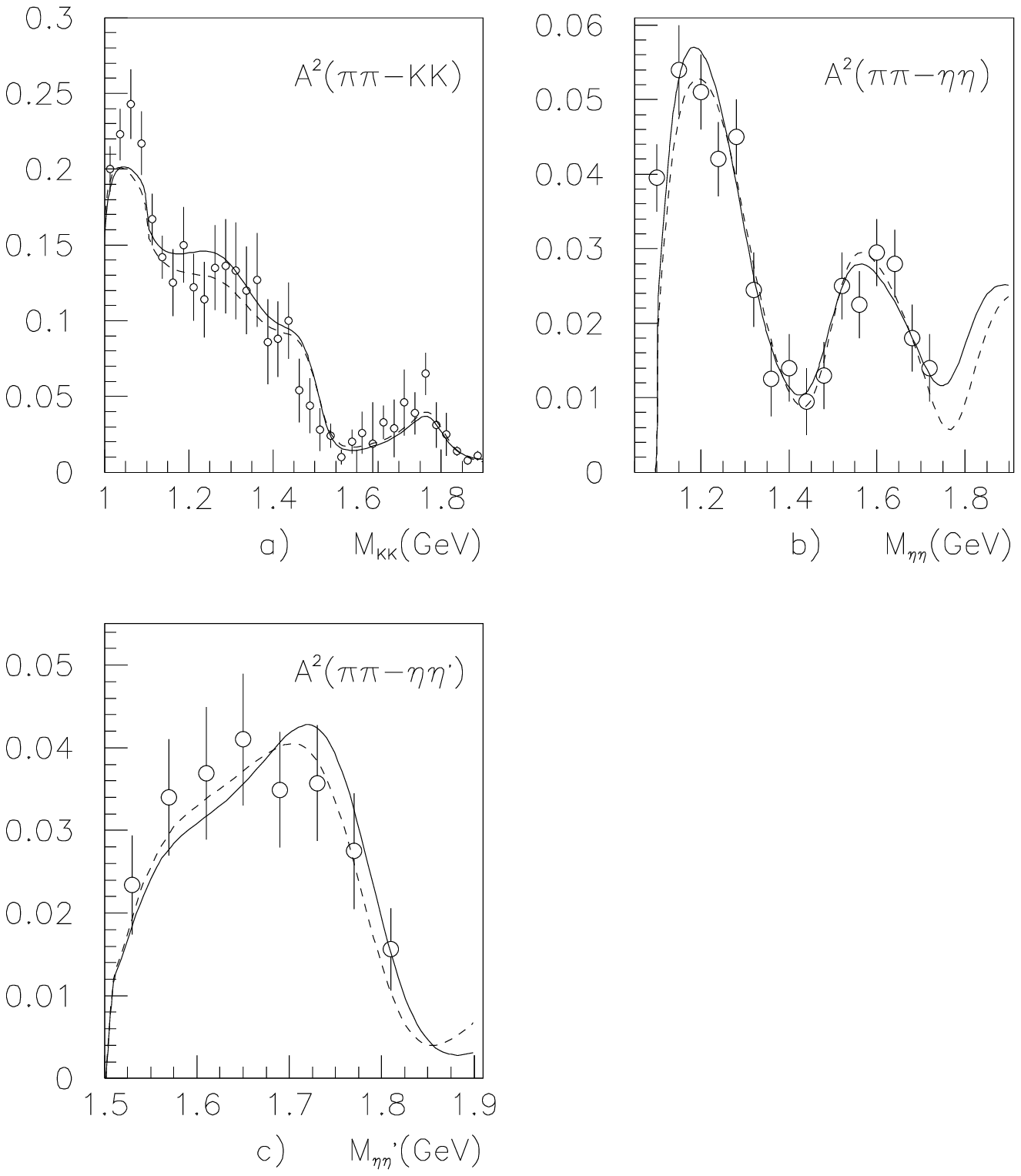,width=16cm}
\caption\protect{The $S$-wave amplitudes squared for transitions:
(a) $\pi\pi\to K\bar K$ \cite{bnl},
(b) $\pi\pi\to \eta\eta$ \cite{gams2} and
(c) $\pi\pi\to \eta\eta'$ \cite{gams3}.
The solid curve corresponds to solution {\bf II-2} and the dashed
curve to  solution {\bf I}.}
\label{fig5}
\end{figure}

\begin{figure}
\epsfig{file=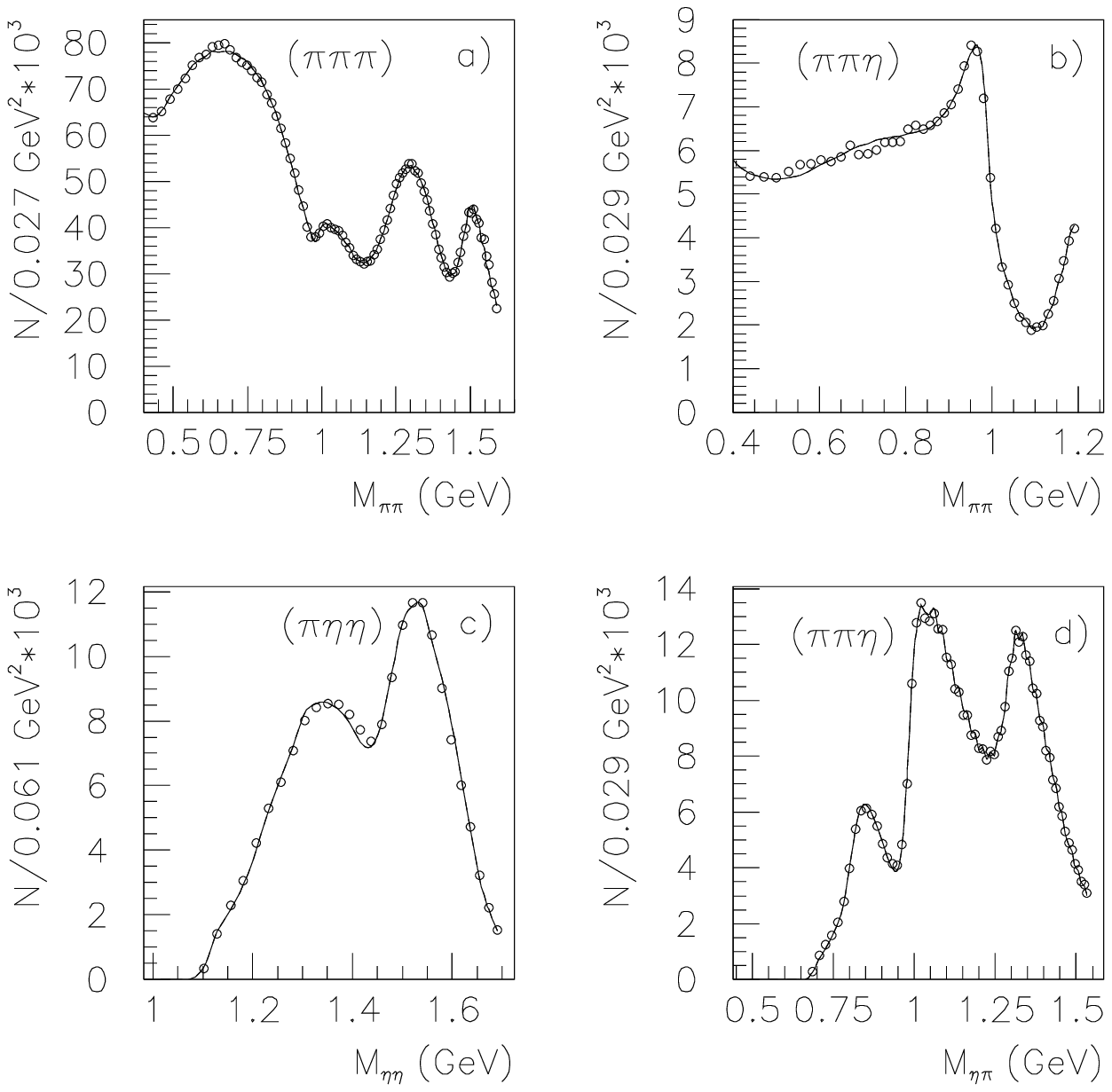,width=16cm}
\caption\protect{ Mass projections of the Dalitz plot on the two-meson
invariant mass for Crystal Barrel data.
The curve corresponds to solution {\bf II-2}.}
\label{fig6}
\end{figure}

\begin{figure}
\epsfig{file=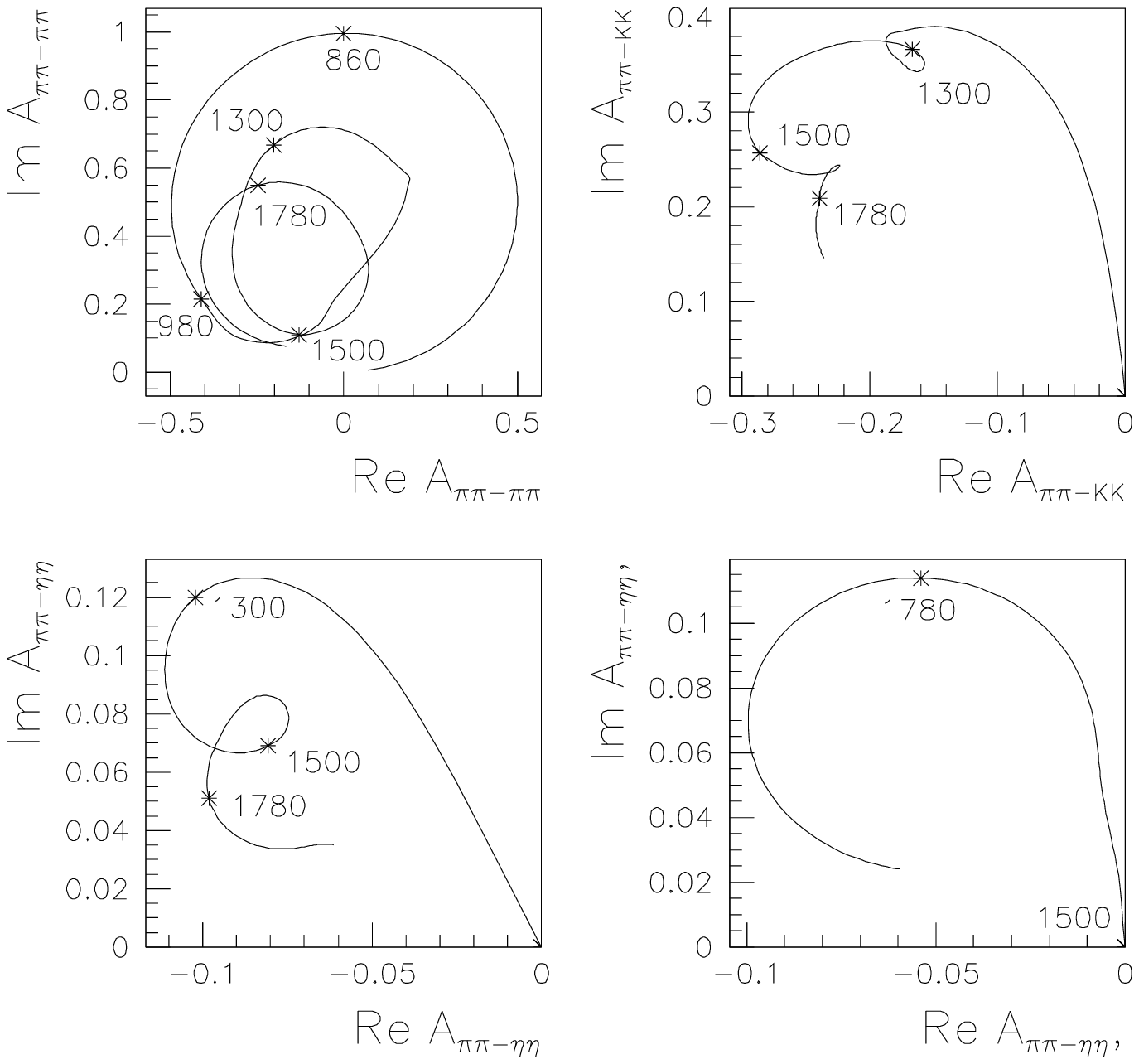,width=15cm}
\caption\protect{ Argand plots for
the $S$-wave scattering amplitudes
in solution {\bf II-2}:  $\pi\pi\to\pi\pi$ (a), $\pi\pi\to K\bar K$
(b), $\pi\pi\to \eta\eta$ (c) and $\pi\pi\to \eta\eta'$ (d).}
\label{fig7}
\end{figure}

\begin{figure}
\epsfig{file=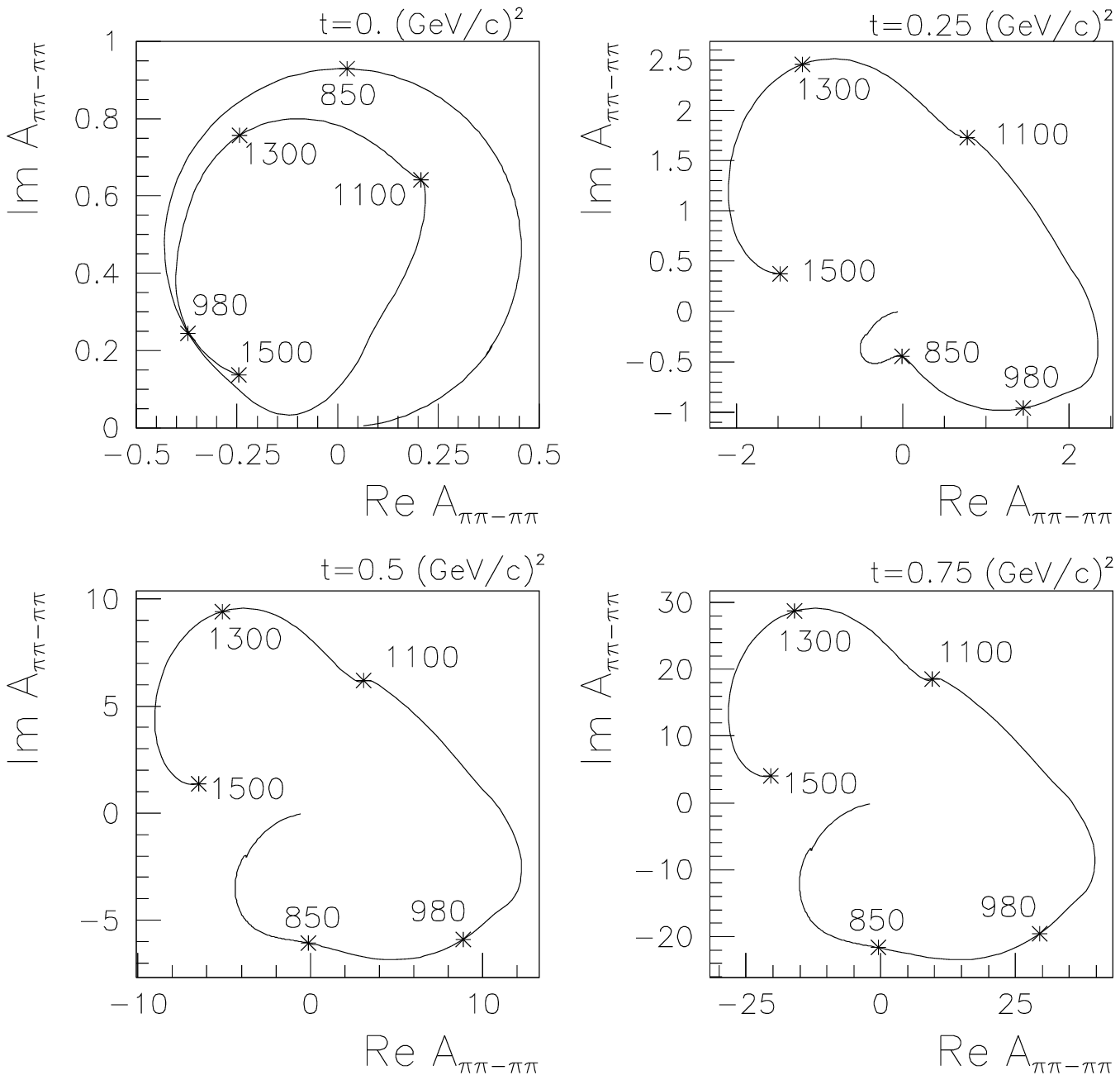,width=15cm}
\caption\protect{Argand plots for the $\pi\pi(t)\to\pi\pi$ S-wave
S-wave scattering amplitudes at different t.}
\label{fig8}
\end{figure}

The scattering amplitude has five
poles in the energy complex plane, four of them  correspond to
relatively narrow resonances while the fifth resonance is very
broad:
\be
f_0(980) \to& (1015\pm 15)- i(43\pm 8)& \; \mbox{MeV},
\nonumber \\
f_0(1300)\to& (1300\pm 20)-i(120\pm 20)&\; \mbox{MeV},
\nonumber \\
f_0(1500)\to& (1499\pm 8)-i(65\pm 10)& \; \mbox{MeV},
\nonumber \\
f_0(1530)\to& (1530^{+90}_{-250}) - i(560\pm 140)&\; \mbox{MeV},
\nonumber \\
f_0(1780)\to& \left \{ \begin{array}{cl}
(1780\pm 30)-i(140\pm 20)\; \mbox{MeV},\\
\qquad (\mbox{Solution}\; I), \\
(1780\pm 50)-i(220\pm 50) \; \mbox{MeV}, \\
\qquad(\mbox{Solution}\; II).
\end{array} \right .
\label{4.2}
\ee
The broad resonance is crucial for the description of the $00^{++}$
wave,
being responsible for large interference effects which are seen in
different reactions. Namely, the resonance $f_0(980)$ reveals itself
as a dip in the S-wave $\pi\pi\to \pi\pi$-spectum, Fig.~\ref{fig2}(a),
 and as a sharp peak in the $\pi\pi (t)\to \pi\pi$ spectra at large
$|t|$, Fig.~\ref{fig3}. The resonance $f_0(1300)$ is seen in the
$\pi\pi (t)\to \pi\pi$ spectra at large $|t|$ as a well shaped bump,
Fig.~\ref{fig3}, while in the $\pi\pi\to \pi\pi$ and $\pi\pi\to
K\bar K$ spectra it reveals itself as a shoulder, Figs.~\ref{fig2} and
~\ref{fig5}.  $f_0(1500)$ is seen as a dip in the $\pi\pi\to\pi\pi$ and
$\pi\pi\to\eta\eta$ spectra, Figs. ~\ref{fig2}, ~\ref{fig5}, and as a
peak in $p\bar p (at \; rest)\to \pi^0\pi^0\pi^0$ reaction,
Fig.~\ref{fig6}.  In all these appearances of $f_0(980)$, $f_0(1300)$
and $f_0(1500)$, their interference with $f_0(1530^{+90}_{-250})$ plays
a decisive role.  In the case of  large interference effects it is
useful to display the amplitude on the Argand-plot. The Argand-plots
for the amplitudes $\pi\pi\to\pi\pi$, $\pi\pi\to\eta\eta$, $\pi\pi\to
K\bar K$, $\pi\pi\to\eta\eta'$ and $\pi\pi(t)\to\pi\pi$ are shown in
Figs. ~\ref{fig7} and ~\ref{fig8}.

Four bare states of Eq. (\ref{4.1}) can be naturally classified as
 nonet partners of the $q\bar q$ multiplets $1^3P_0$ and $2^3P_0$.
The fifth bare state is superfluous for the $q\bar q$ classification
being a good candidate for the lightest scalar glueball. Eq.
(\ref{4.1}) gives two variants for the glueball: either it is a
bare state with mass near 1250 MeV or it is located near 1600 MeV.
Correspondingly, after having imposed the constrains (\ref{25}) and
(\ref{26}), we  found the following variants of the nonet
classification.  For the solution {\bf{I}}:
\begin{description}
\item[I.~~~~]
$f_0^{\mbox{bare}}(720)$ and $f_0^{\mbox{bare}}(1260)$ are $1^3P_0$
nonet partners,\\
$f_0^{\mbox{bare}}(1600)$ and $f_0^{\mbox{bare}}(1810)$ are
$2^3P_0$ nonet partners,\\ $f_0^{\mbox{bare}}(1230)$ is a glueball.
\end{description}

Within solution {\bf{II}}, two variants
describe well the data set:
\begin{description}
\item[II-1.] $f_0^{\mbox{bare}}(720)$ and $f_0^{\mbox{bare}}(1260)$ are $1^3P_0$
nonet partners,\\ $f_0^{\mbox{bare}}(1600)$ and $f_0^{\mbox{bare}}(1810)$ are
$2^3P_0$ nonet partners,  \\ $f_0^{\mbox{bare}}(1230)$ is a glueball;
\item[II-2.] $f_0^{\mbox{bare}}(720)$ and $f_0^{\mbox{bare}}(1260)$ are $1^3P_0$
nonet partners, \\$f_0^{\mbox{bare}}(1230)$ and $f_0^{\mbox{bare}}(1810)$ are
$2^3P_0$ nonet partners,\\ $f_0^{\mbox{bare}}(1600)$ is a glueball.
\end{description}
Tables \ref{table4}-\ref{table5} present parameters which correspond
to these three variants.

Lattice calculations of the gluodynamic glueball \cite{lattglue}
give the mass of the lightest scalar state in the region
1550-1750 MeV that coincides with the variant {\bf{II-2}}.
However, it should be emphasized that the state $f_0^{\mbox{bare}}(1600)$ can
not be identified as a pure gluodynamic glueball because $f_0^{\mbox{bare}}$'s
contain the $q\bar q$-components related to  real parts of the loop
transition diagrams: this problem is discussed in detail in
Refs.  \cite{aas-z,dmatr1,dmatr2}. An extraction of the $q\bar
q$-component from $f^{\mbox{bare}}_0(1600)$ leads to the mass shift of the
state which is not large according to Refs.  \cite{aas-z,dmatr2}:
$f_0^{\mbox{bare}}(1600)\to f_0^{pure\;gluball}(1633)$.

\subsection{$IJ^{PC}=10^{++}$ wave}

Two isovector/scalar resonances are well seen
in the $p\bar p$ annihilation into three
mesons  \cite{km1500,km1900,cbc,bugg,bsz}. The lightest one is
the well known
$a_0(980)$, while the next resonance is the newly discovered
$a_0(1450)$ with mass  $1450\pm 40$  MeV
and width $\Gamma=270\pm 40$ MeV, as  is
given by the Particle Data Group \cite{pdg}.
Let us note that in fitting
the last high statistic Crystal Barrel data with
the T-matrix method used for this wave \cite{km1900,bugg,bsz}
the mass of this
resonance appeared to be a bit higher and equal to
$1520\pm 40$ MeV. A similar result is obtained in the present
K-matrix approach.

For the description of the isovector/isoscalar scattering amplitude, we
use the two-pole $4\times 4$ K-matrix with
two-meson coupling constants given in Table \ref{table2}.

In the first stage of the fit,  the coupling of the
lightest $a_0$-state was allowed to vary in the interval bounded by
$g[f^{\mbox{bare}}_0(720)]$ and $g[f^{\mbox{bare}}_0(1260)]$.  In all the variants of
the fit, the two-meson coupling constant of the lightest state,
$g[a^{\mbox{bare}}_0(lightest \; state)]$, appeared to be very close to the
coupling constant  $g[f^{\mbox{bare}}_0(1260)]$: in the final fit, in line
with  the constraint of Eq. (30), we fix these couplings equal to each
other.  The two-meson coupling of the next isovector/scalar is
fixed to be equal to the couplings of the $2^3P_0$ isoscalar/scalar
states.

\begin{figure}
\epsfig{file=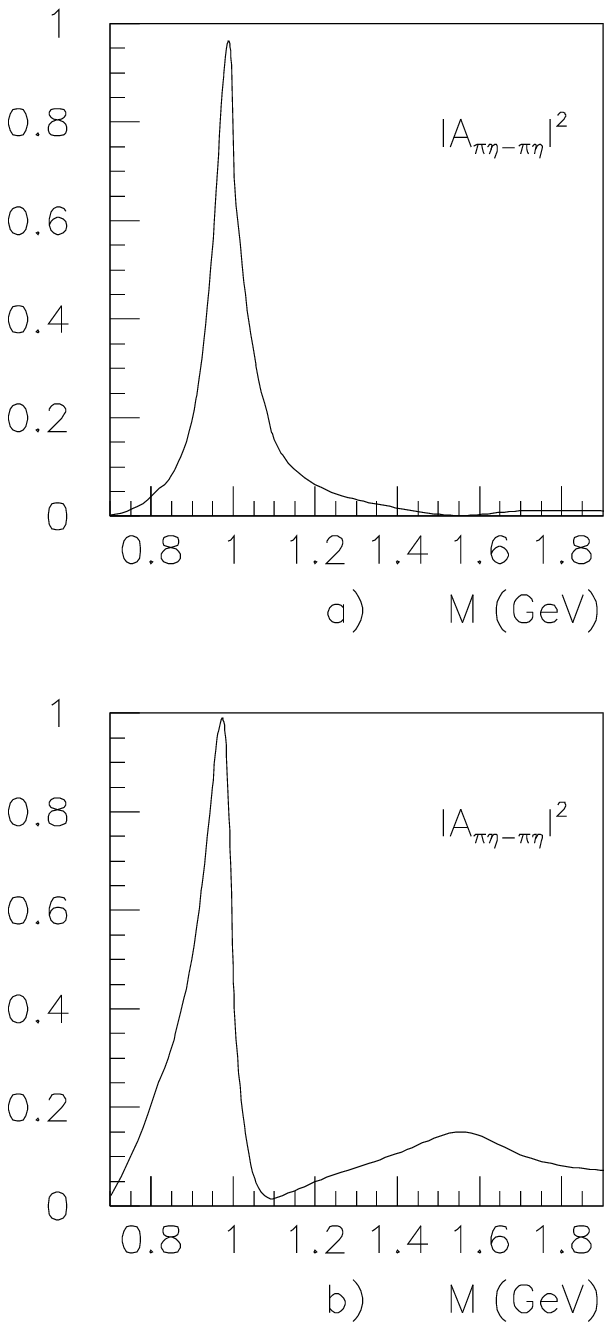,width=11.cm}
\caption\protect{The squared S-wave $\pi\eta\to\pi\eta$ scattering
amplitude: solutions 1 (a) and 2 (b) for the $\pi \eta \to \pi \eta$
scattering amplitude.}
\label{fig9}
\end{figure}

The fit gives two solutions for the $10^{++}$ wave which practically
coincide in terms related to the resonance/bare-state sector and
differ in background terms.
Parameters for the $10^{++}$ wave
and the pole position are given in Table \ref{table7}.
We have for the resonance
positions and the bare states, correspondingly:
\be
a_0(980) &\to \; (988 \pm 6) -i(46\pm 10)\; \mbox{MeV},
\nonumber \\
a_0(1450) &\to \; (1535 \pm 30) -i(146\pm 20)\; \mbox{MeV}
\label{4.4}
\ee
and
\be
a_0^{\mbox{bare}}(964\pm 16)\; , \; \; a_0^{\mbox{bare}}(1670\pm 80) \; .
\label{4.5}
\ee
But these two solutions give different predictions for the scattering
amplitudes: for the first solution (without K-matrix background
terms) the  $\pi\eta \to \pi\eta$
scattering amplitude squared (see Fig. ~\ref{fig9}) has a dip in the
region of $ a_0(1450)$ due to the destructive resonance interference
with the background, while for the second solution
(with the K-matrix background terms) a dip appears at
1100 MeV.  In the present fit, the information on the isovector/scalar
wave comes from  Crystal Barrel data only. These data being rather
sensitive to the pole structure  provide poor information about
K-matrix background terms: it is a source of ambiguities in
our K-matrix solution. But, let us stess, the description of other
partial waves practically does not depend on whether the first or second
solution is used: the variation of parameters is within  the given
errors.

\subsection{$IJ^{PC}=12^{++}$ wave}

Similar to the isovector/scalar case, the $4\times4$ two-pole K-matrix
is used for the description of the $12^{++}$-wave.
Coupling constants of  bare states and the
poles of the scattering amplitude are given in Table \ref{latetable}.
We have determined two bare states:
\be
a_2^{\mbox{bare}}(1314\pm 7),\; \;
a_2^{\mbox{bare}}(1670\pm 75).
\label{4.6}
\ee
The poles of the amplitude are located at
\be
a_2(1320) &\to \; (1309 \pm 6) -i(58\pm 6)\; \mbox{MeV},
\nonumber \\
a_2(1640) &\to \; (1640 \pm 50) -i(122\pm 18)\; \mbox{MeV}
\label{4.7}
\ee
The lightest
state is a well known $a_2(1320)$ resonance, with
mass  $1318\pm 1$ MeV and width $\Gamma=107\pm 5$ MeV,
according to Ref.  \cite{pdg}.

When fitting Crystal Barrel data on the reaction
$p\bar p(at\;rest)\to \eta\pi\pi$, the introduction of the
isovector/tensor resonance with mass 1600-1700
MeV makes an appreciable
improvement of the Dalitz plot description  in this
region.

\subsection{$IJ^{PC}=02^{++}$ wave}

The two lightest  isoscalar/tensor
states, $f_2(1270)$ and $f'_2(1525)$, are well known:
they are members of the nonet $1^3P_2 q\bar q$.
Crystal Barrel data point out  the existence of the resonance
$f_2(1565)$ which helps to describe the $p\bar p\to \pi^0\pi^0\pi^0$
Dalitz plot in the region of large two-pion masses
\cite{cbc,dmatr1,bsz}.
Because of that,  we also begin  our analysis
introducing a three-pole K-matrix amplitude.
However, after imposing the nonet constraints on the $1^3P_2 $
states,  see eqs. (28) and (29),
we obtain that the couplings of the third state
turn out to be negligibly small. Although the description of the
reaction $p\bar p\to \eta\eta\pi^0$ becomes a bit worse
under the imposed constraints
(about 0.1  per degree of freedom for $\chi^2$),
the description of the reaction $p\bar p \to \pi^0\pi^0\pi^0$
(where $f_2(1560)$ is seen
according to \cite{cbc,dmatr1,bsz}) improves  $\chi^2$ by
0.07 giving practically the same total $\chi^2$.

$f_2(1560)$ is not seen in GAMS data; that gives a
strong restriction on the partial width of the resonance decay
into $\pi\pi$ channel, less than 20 MeV.

In our final fit, we have used the two-pole K-matrix
amplitude with the nonet constraints; parameters for this fit are
presented in Table \ref{latetable}.

The K-matrix fit gives the follow\-ing bare iso\-sca\-lar/ten\-sor 
states, the members of the $^3P_2$ nonet:  
\be 
f_2^{\mbox{bare}}(1235\pm 10),\; f_2^{\mbox{bare}}(1530\pm 10),
\nonumber \\
\Phi[f_2^{\mbox{bare}}(1530)]= 86^o \pm 5^o\; .
\label{4.8}
\ee
The K-matrix $02^{++}$
amplitude has poles at the complex mass values:
\be
f_2(1270) \to  (1262 \pm 6) -i(90\pm 7) \mbox{MeV},
\nonumber \\
f'_2(1525) \to  (1518 \pm 9) -i(71 \pm 10 ) \mbox{MeV}.
\label{4.9}
\ee
These values should be compared with masses and half-widths
of Particle Data Group
\cite{pdg} which  are, correspondingly:
$(1275\pm 5)$ MeV, $(92.5\pm 10)$ MeV and $(1525\pm 5)$ MeV,
$(38\pm 5)$ MeV. The width of $f'_2(1525)$ found in our fit
appears to be much larger than one given in PDG. It is quite possible
that in fitting the present data set  we cannot resolve a possible
 D-wave double pole structure in the region of 1530 MeV caused by the
$f'_2(1525)$ and $f_2(1560)$ resonances, for they are located
near the edge of the phase space for Crystal Barrel data, while
GAMS data give a restriction only on the couplings to $\pi\pi$
channel.  It is possible that the additional information from
Crystal Barrel data on $KK\pi$ production together with
 GAMS \cite{gams_f2} and VES data \cite{ves} on $\omega\omega$
production will clarify this point.

\subsection{Nonet classification}

The results of the performed analysis together with the results of the
K-matrix analysis of the $K\pi$ S-wave \cite{alan} allow us to construct
the lightest scalar $q\bar q$ nonet uniquely as
\be
1\;^3P_0:&f_0^{\mbox{bare}}(720\pm 100),& \nonumber \\
&f_0^{\mbox{bare}}(1260\pm 30),&          \nonumber \\
&a_0^{\mbox{bare}}(960\pm 30),&\nonumber \\
&K_0^{\mbox{bare}}(1220^{+50}_{-150}),&\nonumber \\
&\Phi[f^{\mbox{bare}}_0(720)]=&-70^o\;^{+5^o}_{-16^o}\; .
\label{f1}
\ee
The lightest scalar, $f_0^{\mbox{bare}}(720\pm 100)$, is dominantly a
$s\bar s$ state with mixing angle close to the ideal octet one,
$\Phi_{{\rm ideal\; octet}}=-55^o$. The situation with the lightest
scalar nonet is similar to that with the lightest pseudoscalar nonet,
where the mixing angle for the $\eta$-meson is also close to the
$\Phi_{{\rm ideal\; octet}}$:  this definitely  indicates  the
degeneration of the lightest $00^{++}$ and $00^{-+}$ states.

The multiplet of the lightest tensor states appears as
\be
1\;^3P_2:&f_2^{\mbox{bare}}(1240\pm 10),&     \nonumber \\
&  f_2^{\mbox{bare}}(1522\pm 10),&     \nonumber \\
& a_2^{\mbox{bare}}(1311\pm 3),&       \nonumber \\
&K_2^{*}(1430)&                        \nonumber \\
&\Phi[f_2^{\mbox{bare}}(1240)]=&-10^o\pm 3^o\; .
\label{f2}
\ee
The K-matrix analysis of the $\pi K$ $D$-wave is not done:
the $(J=2)\pi K$ resonance with mass $1431\pm 3$ is reported in
Ref.  \cite{aston}; we have used this resonance to complete the
multiplet (40).

Our analysis gives two variants for the $2\;^3P_0$
$q\bar q$-nonet:\\
First variant:\\
\be
2\;^3P_0:&f_0^{\mbox{bare}}(1600\pm 50),& \nonumber \\
& f_0^{\mbox{bare}}(1810\pm 30),&  \nonumber \\
&a_0^{\mbox{bare}}(1650\pm 50),& \nonumber \\
& K_0^{\mbox{bare}}(1885^{+50}_{-100}),&\nonumber \\
&\Phi[f^{\mbox{bare}}_0(1810)]=&84^o\pm 5^o
\ee
The state $K_0^{\mbox{bare}}(1885^{+50}_{-150})$ is fixed by the analysis
\cite{alan} of the $K\pi$ $S$-wave. In this variant the lightest
glueball state is $f_0^{\mbox{bare}}(1230^{+150}_{-30})$. In the second
variant the lightest glueball state is identified as
$f_0^{\mbox{bare}}(1600)$, namely:\\
\be
2\;^3P_0:&f_0^{\mbox{bare}}(1230^{+150}_{-30}), &\nonumber \\
&f_0^{\mbox{bare}}(1810\pm 30),&       \nonumber \\
&a_0^{\mbox{bare}}(1650\pm 50), &       \nonumber \\
&K_0^{\mbox{bare}}(1885^{+50}_{-100}),&\nonumber \\
& \Phi[f^{\mbox{bare}}_0(1810)]=&44^o\pm 10^o
\ee

\section{Resonances \fres
and  \ares: are they $K\bar K$ molecules?}

First, let us discuss the origin of $f_0(980)$.
GAMS data for the $f_0(980)$ production at large
$|t|$, see Fig.~\ref{fig3}, directly demonstrate
that this resonance has a hard component, while the location of the pole
near the $K\bar K$ threshold definitely says that its kaon component is
a long-range one. The existence of the long-range component gives rise to
discussion about molecular structure for this state \cite{kk}.
The problem  to discuss is how substantial are these components  in
the formation of the resonance. Remember that the short-range
component (with $ r < 1$ fm) is a subject of quark/gluon considerations
and quark systematics.

The resonance $f_0(980)$ corresponds to the two poles   located at
(in MeV units):
\be
M=1015-i46\;  &(II \; \mbox{sheet}, \;\;\mbox{under}\;
\pi \pi -\mbox{cut})\,,
\nonumber \\
M=936-i238  &(III\; \mbox{sheet}, \;\; \mbox{under}\; \pi \pi \;
\mbox{and} \; K\bar K \;\mbox{cuts}).
\nonumber
\ee
The second pole appears due to well-known double-pole
structure caused by the
$K \bar K$ -threshold (see, for example, \cite{flatte}), while the
first pole at $M=1015-i46$ MeV generates the leading irregularities in
$\pi \pi$ spectra.

The restored K-matrix amplitude allows one to see the role of
the $K\bar K$
component in the formation of $f_0(980)$, thus clarifying if this
resonance is a descendant of a $q\bar q$ state or is
a molecular-type system. To this aim, let us switch off the $f_0(980)$
decay processes (transitions into $\pi\pi$ and $K\bar K$) and look at
the dynamics of pole positions, with gradual onset of
couplings.  For the gradual change of couplings
we performed the  replacement in
the K-matrix $00^{++}$-amplitude:
\be
 g^{(\alpha)}_{a} \to \xi
g^{(\alpha)}_{a} \; .
\ee
 Parameter $\xi $  varies in the interval:
\be
 0 < \xi
\leq 1  \; .
\ee
At $\xi \to 0$ the decay channels are switched off, and
we have a bare state,  while at $\xi =1$ the real case is restored.

\begin{figure}
\epsfig{file=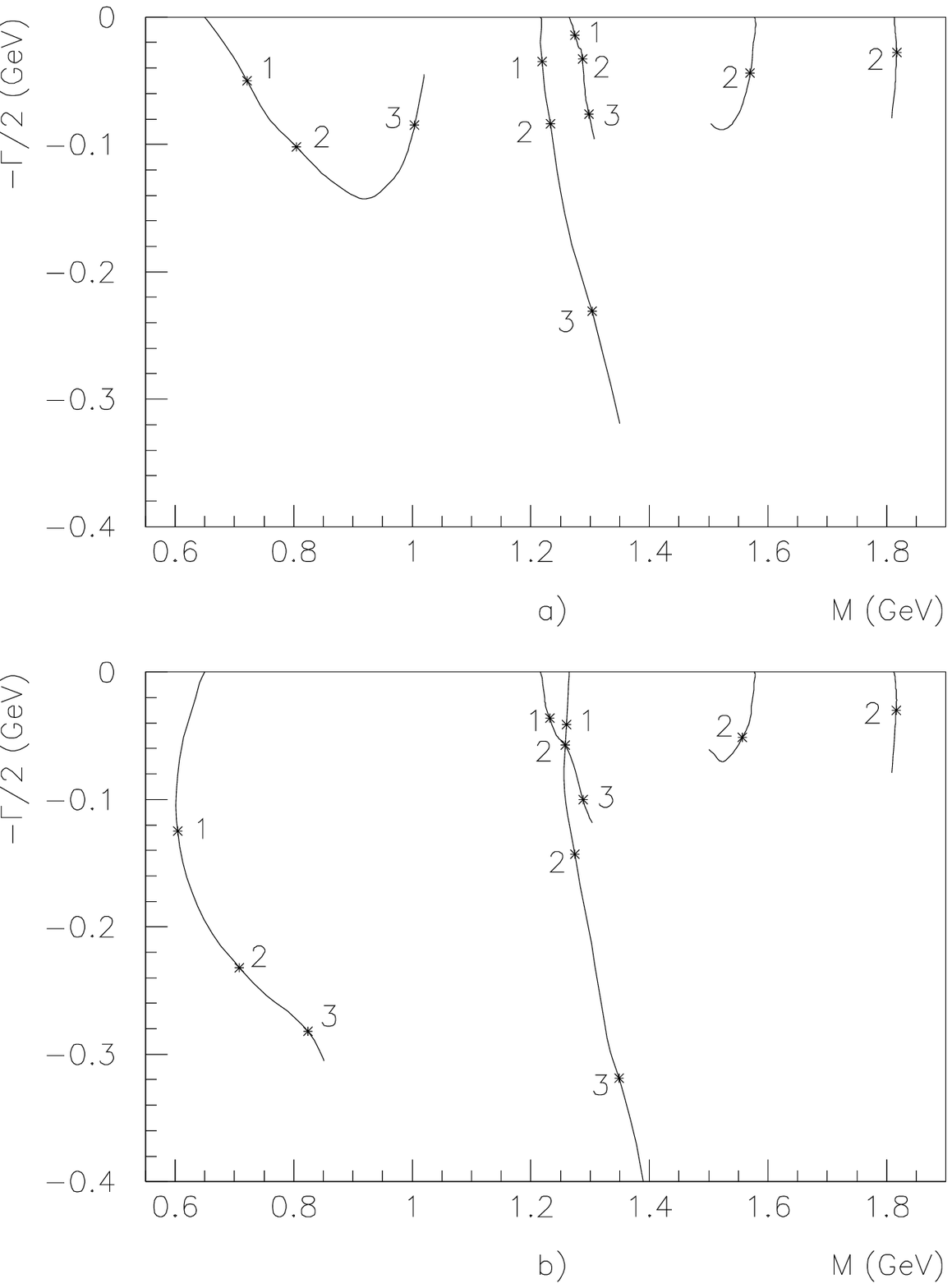,width=15cm}
\caption\protect{Location of $00^{++}$ amplitude poles
in the complex-$\sqrt s$ plane
$(M=Re\sqrt s,\;\;-\Gamma/2=Im\sqrt s)$
after replacing $g_a^{\alpha} \to \xi g_a^{\alpha}$,
on the sheet under the $\pi\pi$ cut (a) and
on the sheet under $\pi\pi$ and $K\bar K$ cuts (b).
The case $\xi\to 0$
gives the positions of masses of bare states;
$\xi=1$ corresponds to the real case. The point 1 corresponds to
$\xi=0.4$, 2 to $\xi=0.6$ and 3 to $\xi=0.9$.}
\label{latefigure}
\end{figure}

At  $\xi \to 0$ the masses of the lightest scalar bare states are 650
and 1260 MeV (the positions of the K-matrix poles).
The trajectories of states with increasing $\xi$
are shown in Fig.~\ref{latefigure}.

The crucial point is what component,  $\pi\pi$ or $K\bar K$, is mainly
responsible for the mass shift  from 650 MeV to $1020-i48$ MeV.
We can elucidate this point, switching off the $K\bar K$
component and leaving $\pi \pi$  untouched, and vice versa.
In the first case the mass of $f_0(980)$ state is:
\be
M(\mbox{without} \; K\bar K)=974-i115 \; \mbox{MeV} \; .
\ee
One sees that the mass shift
\be
\delta_{K\bar K} =&M(\xi =1) -M(\mbox{without} \; K\bar K) =
\nonumber \\
& 41+i67 \; \mbox{MeV}
\ee
is not large:
the $K\bar K$-component which is responsible for the value of
$\delta$ does not play an important role in the formation of  the mass
of $f_0(980)$. In the second case, when the $\pi\pi$ component is
switched off, we
get the nearest state to the $K\bar K$ threshold, which is located  at:
\be
M(\mbox{without} \;
\pi\pi)=810-i10 \; \mbox{MeV} \; .
\ee
So the mass shift is
\be
\delta_{\pi\pi} =&M(\xi=1) -M(\mbox{without} \; \pi\pi) =
\nonumber \\
205+i36 \; \mbox{MeV} ,
\ee
being much larger  than  $\delta_{K\bar K}$.
The  transition into real pions,
\be
f_0^{\mbox{bare}}(720) \to \pi \pi
\ee
is mainly
in  charge of the mixing of the lightest scalar $q\bar q$ state
with other scalars thus shifting  its mass by chance to the region
of the next threshold, $K\bar K$. The $K\bar K$ component  of
$f_0(980)$ is of the molecule-type: relative kaon momenta  are
small, so relative distances are large. But, let us stress again,
the two-kaon component does not play a crucial role in the
formation of the mass of $f_0(980)$.

\section{Conclusion}

We have performed the K-matrix anlysis of  GAMS data on the S-
and D-wave $\pi^0 \pi^0$, $\eta \eta$ and $\eta \eta '$ data, together
with  data obtained by BNL and Crystal Barrel Collaboration. Partial
amplitudes for the states $00^{++}$, $02^{++}$, $10^{++}$ and
$12^{++}$ are
investigated in the mass region up to 2000 MeV, the poles of these
amplitudes are found, see Tables \ref{table4}-\ref{latetable}. Pole terms
of the K-matrix are restored, i.e. the bare states with quantum numbers
$00^{++}$, $02^{++}$, $10^{++}$ and $12^{++}$ are found. The quark
content of these bare states is determined, based on the relations
between coupling constants of the decays: this allows to restore the
quark nonets $1^3P_0$, $2^3P_0$ and $1^3P_2$. The performed analysis
confirms the result of Ref.  [6] which is based on the K-matrix
analysis of the $00^{++}$ wave only: in the region 1200-1600 MeV there
exists a scalar/isoscalar state which is superfluous for the $q\bar q$
systematics. This state is a good candidate for the lightest scalar
glueball.

The analysis indicates  the degeneration of the lightest $00^{++}$ and
$00^{-+}$ states.

{\bf Acknowledgements}

We thank D. V. Bugg,  S. S. Gershtein,
A. K. Likhoded, L. Montanet and B. S. Zou for useful discussions and
Crystal Barrel Collaboration for providing us with the data.  VVA and
AVS are grateful to the RFFI (Grant 96-02-17934) and INTAS-RFBR (Grant
95-0267) for  financial support.

\onecolumn

\newpage
\widetext

\begin{table}
\caption\protect{Coupling constants given by quark combinatorics for
a $q\bar q$-meson
decaying into two pseudoscalar mesons in the leading terms
of the $1/N$ expansion
and for glueball decay in the next-to-leading terms
of the $1/N$ expansion.
$\Phi$ is the mixing angle for $n\bar n$ and $s\bar s$ states, and
$\Theta$ is the mixing angle for $\eta -\eta'$ mesons:
$\eta=n\bar n \cos\Theta-s\bar s \sin\Theta$ and
$\eta'=n\bar n \sin\Theta+s\bar s \cos\Theta$.
Glueball decay couplings in the leading terms of $1/N$ expansion
are obtained by the replacements $g/\sqrt 2\cos\Phi\to G^L$,
$g\sin\Phi\to\sqrt\lambda G^L$.}
\label{table1}
\begin{tabular}{|c|c|c|c|}
~      &     ~                    &  ~                     &~           \\
~      & The $q\bar q$-meson decay&Glueball decay couplings&Identity   \\
~      & couplings in the         &in the next-to-      &   factor in  \\
Channel& leading terms of $1/N$   &leading terms of $1/N$ &phase space \\
~      & expansion
&expansion      &~   \\
~      &     ~                    &  ~                     &~           \\
\hline
~ & ~ & ~ & ~ \\
$\pi^0\pi^0$ &
$g\;\cos\Phi/\sqrt{2}$& 0 & 1/2  \\
~ & ~ & ~ & ~ \\
$\pi^+\pi^-$ & $g\;\cos\Phi/\sqrt{2}$ & 0 &  1   \\
~ & ~ & ~ & ~ \\
$K^+K^-$ & $g (\sqrt 2\sin\Phi+\sqrt \lambda\cos\Phi)/\sqrt 8 $ & 0 & 1 \\
~ & ~ & ~ & ~ \\
$K^0K^0$ & $g (\sqrt 2\sin\Phi+\sqrt \lambda\cos\Phi)/\sqrt 8 $ & 0 & 1 \\
~ & ~ & ~ & ~ \\
$\eta\eta$ &
$g\left (\cos^2\Theta\;\cos\Phi/\sqrt 2 +
\sqrt{\lambda}\;\sin\Phi\;\sin^2\Theta\right )$
&$2g_G(\cos\Theta-\sqrt{\frac{\lambda}{2}}\sin\Theta )^2$ &
1/2 \\
~ & ~ & ~ & ~ \\
$\eta\eta'$ &
$g\sin\Theta\;\cos\Theta\left(\cos\Phi/\sqrt 2-
\sqrt{\lambda}\;\sin\Phi\right ) $ 
&$2g_G(\cos\Theta-\sqrt{\frac{\lambda}{2}}\sin\Theta)
(\sin\Theta+\sqrt{\frac{\lambda}{2}}\cos\Theta)$
 & 1\\
~ & ~ & ~ & ~ \\
$\eta'\eta'$ &
$g\left(\sin^2\Theta\;\cos\Phi/\sqrt 2+
\sqrt{\lambda}\;\sin\Phi\;\cos^2\Theta\right)$ 
&$2g_G (\sin\Theta+\sqrt{\frac{\lambda}{2}}\cos\Theta )^2$ &
1/2 \\
~ & ~ & ~ & ~ \\
\end{tabular}
\end{table}

\begin{table}
\caption\protect{Coupling constants given by quark combinatorics for scalar
mesons $K^{-}_{0}$  and $a^{-}_{0}$ decaying into two
 pseudoscalar mesons in the leading terms
of the $1/N$ expansion.}
\label{table2}
\begin{tabular}{|c|c|c|c|}
~      &     ~                    &  ~        &~           \\
Channel & The $s\bar u$- meson     & Channel   &  The $d\bar u$- meson \\
~      & decay couplings        &  ~          & decay couplings \\
~      &     ~                    &  ~        &~           \\
\hline ~ & ~ & ~ & ~ \\
$\bar K^0\pi^{-}$ & $g\;(-)\frac 12$
& $\eta \pi^{-}$ & $g\;\frac{1}{\sqrt{2}}\cos\Theta $ \\
~ & ~ & ~ & ~ \\
$ K^{-}\pi^{0}$ & $g\;\frac{1}{\sqrt{8}}$
& $\eta' \pi^{-}$ & $g\;\frac{1}{\sqrt{2}}\sin\Theta $ \\
~ & ~ & ~ & ~ \\
$ K^{-}\eta$ & $g\;\frac{1}{\sqrt{8}}(\cos\Theta
-\sqrt{2\lambda}\sin\Theta)$ & $K^{0}K^{-}$ &
$g\;\frac {\sqrt\lambda}{2} $ \\
~ & ~ & ~ & ~ \\
$ K^{-}\eta'$ & $g\;\frac{1}{\sqrt 8}(\sin\Theta
+\sqrt{2\lambda}\cos\Theta)$ & - & - \\
~ & ~ & ~ & ~ \\
\end{tabular}
\end{table}

\begin{table}
\caption\protect{$\chi^2$ values for the K-matrix solutions.
\label{table3}}
\begin{tabular}{|l|c|c|c|c|}
~ &  Solution I& Solution II-1 & solution II-2& Number of points\\
\hline
Crystal Barrel 
data \cite{cbc}& ~ &~ & ~ & ~ \\
$p\bar p\to \pi^0\pi^0\pi^0$   & 1.52  & 1.41  & 1.42 & 1338\\
$p\bar p\to \pi^0\eta\eta$     & 1.57  & 1.60  & 1.59 & 1798\\
$p\bar p\to \pi^0\pi^0\eta$    & 1.38  & 1.43  & 1.43 & 1738\\
\hline
$\pi^+\pi^-\to\pi^0 \pi^0$ &~   &~ & ~ &~ \\
S-wave 
GAMS data \cite{gams1}        & 1.47 &  1.71 &  1.59 &   70 \\
D-wave 
GAMS data \cite{gams1}        & 1.63 &  2.16 &  2.14 &   56 \\
SD-correlator \cite{gams1}        & 1.82 &  2.26 &  2.12 &   47 \\
\hline
t-dependent
GAMS data \cite{gams1} & ~ &~ & ~ &~ \\
$ 0.00<|t|<0.20$            & 3.03  & 3.42  & 3.37  & 21  \\
$ 0.30<|t|<1.00$            & 2.64  & 3.25  & 2.98  & 38  \\
$ 0.35<|t|<1.00$            & 1.30  & 1.55  & 1.44  & 38  \\
$ 0.40<|t|<1.00$            & 2.75  & 2.48  & 2.79  & 38  \\
$ 0.45<|t|<1.00$            & 1.92  & 1.49  & 1.67  & 38  \\
$ 0.50<|t|<1.00$            & 2.29  & 1.85  & 2.04  & 38  \\
\hline
GAMS data \cite{gams2,gams3} & ~  &~ & ~ &~ \\
$\pi\pi\to \eta\eta$   & 0.70  & 0.97 & 0.87  & 16 \\
$\pi\pi\to \eta\eta'$  & 0.49  & 0.65 & 0.64  & 8  \\
\hline
$\pi\pi\to K\bar K$ &  ~   &  ~    & ~    & ~   \\
BNL data \cite{bnl}              & 0.88 &  0.77 & 0.97 &  35 \\
\end{tabular}
\end{table}

\begin{table}
\caption\protect{Masses, coupling constants (in GeV) and mixing angles
(in degrees)
for the $f_0^{\mbox{bare}}$-resonances for solution I.
The errors reflect the boundaries for a satisfactory
description of the data. II sheet is under the $\pi\pi$ and $4\pi$
cuts; IV sheet is under the $\pi\pi$, $4\pi$, $K\bar K$ and $\eta \eta$
cuts; V sheet is under the $\pi\pi$, $4\pi$, $K\bar K$, $\eta \eta$ and
$\eta \eta'$ cuts.}
\label{table4}
\begin{tabular}{|l|ccccc|}
~& ~ &\multicolumn{3}{c}{Solution I-1} &~  \\
\hline
~ & $\alpha=1$ &$\alpha=2$ & $\alpha=3$ & $\alpha=4$ & $\alpha=5$ \\
\hline
~         &~ & ~ & ~ & ~ & ~ \\
M              & $0.651^{+.120}_{-.030}$ &$1.247^{+.150}_{-.030}$ &
$1.253^{+.015}_{-.045}$ & $1.684^{+.010}_{-.045}$ &
$1.792^{+.040}_{-.040}$ \\
~         &~ & ~ & ~ & ~ & ~ \\
$g^{(\alpha)}$ &$1.318^{+.100}_{-.100}$ &$0.597^{+.050}_{-.100}$
&$0.879^{+.080}_{-.050}$ &
$0.702^{+.020}_{-.060}$ &$0.702^{+.020}_{-.060}$\\
~         &~ & ~ & ~ & ~ & ~ \\
$g_G$     &0 &$-0.135^{+.050}_{-.050}$ &  0 & 0 & 0 \\
~         &~ & ~ & ~ & ~ & ~ \\
$g_{5}^{(\alpha)}$& 0 &$0.944^{+.100}_{-.150}$ & 0 &
$0.898^{+.070}_{-.150}$ & $0.302^{+.150}_{-.070}$  \\
~         &~ & ~ & ~ & ~ & ~ \\
$\Phi_\alpha $  & -(71.5$^{+3}_{-15}$)
&21.5$^{+8}_{-8}$ & 14.1$^{+10}_{-5}$ & -6.0$^{+10}_{-10}$
& 89$^{+5}_{-15}$\\
~         &~ & ~ & ~ & ~ & ~ \\
\hline
~ & $a=\pi\pi$ &$a=K\bar K$ & $a=\eta\eta$ & $a=\eta\eta'$ & $a=4\pi$ \\
\hline
~         &~ & ~ & ~ & ~ & ~ \\
$f_{1a} $ &$0.455^{+.100}_{-.100}$ &$ 0.061^{+.100}_{-.100}$ &
$0.501^{+.100}_{-.100}$ &$0.448^{+.100}_{-.100}$ &
$-0.129^{+.060}_{-.060}$ \\
\hline
~ &  ~ & $f_{ba}=0$ &$b=2,3,4,5$ & ~ & ~ \\
%~         &~ & ~ & ~ & ~ & ~ \\
~ & \multicolumn{2}{c}{$g^{(1)}_3=-0.259^{+0.045}_{-0.045}$} &
\multicolumn{2}{c}{$g^{(1)}_4=-0.275^{+0.100}_{-0.100}$} &
$s_0=3.25^{+\infty}_{-1.0}$ \\
\hline
~ & ~&\multicolumn{3}{c}{Pole position}& ~  \\
II sheet     & $1.006^{+.008}_{-.008}$& ~ & ~ & ~ & ~ \\
 ~ & $-i(0.048^{+.008}_{-.008})$ & ~ & ~ & ~ & ~\\
\hline
IV sheet       & ~ &$1.303^{+.010}_{-.020}$&$1.496^{+.008}_{-.004}$
              &$1.670^{+.100}_{-.150}$ & ~ \\
~     & ~ &$-i(0.138^{+.015}_{-.025})$&$-i(0.059^{+.005}_{-.005})$
              &$-i(0.760^{+.080}_{-.170})$ & ~ \\
\hline
V sheet    & ~ & ~ & ~ & ~ & $1.775^{+.015}_{-.015}$ \\
~       & ~ & ~ & ~ & ~  & $-i(0.056^{+.015}_{-.010})$   \\
\end{tabular}
\end{table}

%----------------------------------------------------------------
%\newpage
\begin{table}
\caption\protect{Masses, coupling constants (in GeV) and mixing angles (in
degrees) for the $f_0^{\mbox{bare}}$-resonances for solutions II-1
and II-2.}
\label{table5}
\begin{tabular}{|l|ccccc|}
\hline
~ & ~ &\multicolumn{3}{c}{Solution II-1}& ~  \\
\hline
~ & $\alpha=1$ &$\alpha=2$ & $\alpha=3$ & $\alpha=4$ & $\alpha=5$ \\
\hline
~         &~ & ~ & ~ & ~ & ~ \\
M              & $0.651^{+.120}_{-.030}$ &$1.246^{+.150}_{-.035}$ &
$1.263^{+.015}_{-.045}$ & $1.595^{+.030}_{-.040}$ &
$1.832^{+.030}_{-.050}$ \\
~         &~ & ~ & ~ & ~ & ~ \\
$g^{(\alpha)}$ &$1.385^{+.100}_{-.100}$ &$0.375^{+.070}_{-.050}$
&$0.923^{+.080}_{-.050}$ &
$0.424^{+.050}_{-.050}$ &$0.424^{+.070}_{-.050}$\\
~         &~ & ~ & ~ & ~ & ~ \\
$g_G$     &0 &$-0.017^{+.050}_{-.050}$ &  0 & 0 & 0 \\
~         &~ & ~ & ~ & ~ & ~ \\
$g_{5}^{(\alpha)}$& 0 &$0.705^{+.100}_{-.100}$ & 0 &
$0.552^{+.070}_{-.070}$ & $-0.557^{+.070}_{-.070}$  \\
~         &~ & ~ & ~ & ~ & ~ \\
$\Phi_\alpha $  & -(70.1$^{+3}_{-15}$)
&30.0$^{+8}_{-8}$ & 18.3$^{+8}_{-5}$ & 20.6$^{+08}_{-15}$
&-64.4$^{+10}_{-10}$\\
~         &~ & ~ & ~ & ~ & ~ \\
\hline
~ & $a=\pi\pi$ &$a=K\bar K$ & $a=\eta\eta$ & $a=\eta\eta'$ & $a=4\pi$ \\
\hline
~         &~ & ~ & ~ & ~ & ~ \\
$f_{1a} $ &$0.440^{+.100}_{-.100}$ &$-0.064^{+.100}_{-.100}$ &
$0.387^{+.100}_{-.100}$ &$0.419^{+.100}_{-.100}$ &
$-0.165^{+.060}_{-.060}$ \\
\hline
~ &  ~ & $f_{ba}=0$ & $b=2,3,4,5$ & ~& ~\\
~         &~ & ~ & ~ & ~ & ~ \\
~ & \multicolumn{2}{c}{$g^{(1)}_3=-0.239^{+0.045}_{-0.045}$} &
\multicolumn{2}{c}{$g^{(1)}_4=-0.284^{+0.100}_{-0.100}$} &
$s_0=3.28^{+\infty}_{-1.0}$ \\
\hline
~ & ~ & \multicolumn{3}{c}{Pole position}& ~  \\
II sheet      &$1.017^{+.008}_{-.008}$ & ~ & ~ & ~ & ~\\
  ~           &$-i(0.049^{+.008}_{-.008})$ & ~ & ~ & ~ & ~\\
\hline
IV sheet  & ~ &$1.311^{+.010}_{-.020}$ & $1.500^{+.004}_{-.006}$
              &$1.470^{+.150}_{-.100}$& ~ \\
 ~        & ~ &$-i(0.117^{+.015}_{-.025})$ & $-i(0.063^{+.003}_{-.006})$
              &$-i(0.545^{+.080}_{-.080})$&~ \\
\hline
V sheet   & ~ & ~ & ~ & ~ &$1.814^{+.015}_{-.015}$ \\
 ~        & ~ & ~ & ~ & ~  &$-i(0.082^{+.025}_{-.010})$  \\
\hline
%\end{tabular}
%\end{table}
%%----------------------------------------------------------------
%\begin{table}
%\caption\protect{Masses, coupling constants (in GeV) and mixing angles
%(in
%degrees) for the $f_0^{\mbox{bare}}$-resonances for solution II-2.}
%\label{table6}
%\begin{tabular}{|l|ccccc|}
\hline
~ & ~ & \multicolumn{3}{c}{Solution II-2} & ~ \\
\hline
~ & $\alpha=1$ &$\alpha=2$ & $\alpha=3$ & $\alpha=4$ & $\alpha=5$ \\
\hline
~         &~ & ~ & ~ & ~ & ~ \\
M              & $0.651^{+.120}_{-.030}$ &$1.219^{+.150}_{-.030}$ &
$1.267^{+.015}_{-.045}$ & $1.584^{+.010}_{-.045}$ &
$1.817^{+.040}_{-.040}$ \\
~         &~ & ~ & ~ & ~ & ~ \\
$g^{(\alpha)}$ &$1.351^{+.100}_{-.100}$ &$0.435^{+.070}_{-.050}$
&$0.901^{+.080}_{-.050}$ &
$0.433^{+.050}_{-.050}$ &$0.435^{+.070}_{-.050}$\\
~         &~ & ~ & ~ & ~ & ~ \\
$g_G$     &0  &  0 & 0 &$-0.005^{+.050}_{-.050}$& 0 \\
~         &~ & ~ & ~ & ~ & ~ \\
$g_{5}^{(\alpha)}$& 0 &$0.719^{+.100}_{-.100}$ & 0 &
$0.542^{+.070}_{-.070}$ & $-0.512^{+.070}_{-.070}$  \\
~         &~ & ~ & ~ & ~ & ~ \\
$\Phi_\alpha $  & -(69.5$^{+3}_{-15}$)
&40.7$^{+8}_{-8}$ & 19.6$^{+10}_{-5}$
& 20.0$^{+08}_{-15}$ &-54$^{+10}_{-10}$\\
~         &~ & ~ & ~ & ~ & ~ \\
\hline
~ & $a=\pi\pi$ &$a=K\bar K$ & $a=\eta\eta$ & $a=\eta\eta'$ & $a=4\pi$ \\
\hline
~         &~ & ~ & ~ & ~ & ~ \\
$f_{1a} $ &$0.459^{+.100}_{-.100}$ &$ 0.046^{+.100}_{-.100}$ &
$0.405^{+.100}_{-.100}$ &$0.420^{+.100}_{-.100}$ &
$-0.214^{+.060}_{-.060}$ \\
\hline
~ &  ~ & $f_{ba}=0$ & $b=2,3,4,5$&~ & ~\\
~         &~ & ~ & ~ & ~ & ~ \\
~ & \multicolumn{2}{c}{$g^{(1)}_3=-0.241^{+0.045}_{-0.045}$} &
\multicolumn{2}{c}{$g^{(1)}_4=-0.273^{+0.100}_{-0.100}$} &
$s_0=3.05^{+\infty}_{-1.0}$ \\
\hline
~ & ~ &\multicolumn{3}{c}{Pole position}& ~  \\
II sheet      &$1.020^{+.008}_{-.008}$& ~&~&~&~ \\
  ~           &$-i(0.048^{+.008}_{-.008})$& ~&~&~&~\\
\hline
IV sheet   &~ &$1.304^{+.010}_{-.020}$ & $1.505^{+.004}_{-.008}$
              &$1.420^{+.150}_{-.070}$ & ~ \\
  ~        & ~&$-i(0.118^{+.015}_{-.025})$ & $-i(0.063^{+.003}_{-.006})$
              &$-i(0.540^{+.080}_{-.080})$ & ~ \\
\hline
V sheet   & ~ & ~ & ~ & ~     & $1.809^{+.015}_{-.015}$ \\
  ~       & ~ & ~ & ~ & ~     & $-i(0.080^{+.025}_{-.010})$   \\
\hline
\end{tabular}
\end{table}
%----------------------------------------------------------------
%\newpage
\begin{table}
\caption\protect{Masses and coupling constants (in GeV) for
$a_0$ resonances. The star denotes that the parameter is fixed.}
\label{table7}
\begin{tabular}{|l|cc|cc|}
\hline
~& \multicolumn{3}{c}{~} & ~\\
~&\multicolumn{3}{c}{$a_0$-resonances without K-matrix background
term} &~ \\
\hline
~ & Solution I-1 & ~
& Solutions II-(1,2)& ~  \\
\hline
~ & $\alpha=1$ &$\alpha=2$  & $\alpha=1$ &$\alpha=2$   \\
\hline
~         &~ & ~ & ~ & \\
M   &
$0.963^{+.015}_{-.015}$ &$1.630^{+.100}_{-.040}$ &
$0.965^{+.015}_{-.015}$ &$1.654^{+.100}_{-.040}$ \\
~         &~ & ~ & ~ & \\
$g^{(\alpha)}$ &
$0.879^{+.100}_{-.100}$ &$ 0.702^*        $ &
$0.901^{+.100}_{-.100}$ &$ 0.435^*       $ \\
~         &~ & ~ & ~ & \\
$g_{4}^{(\alpha)}$&
$ 0.598^{+.150}_{-.050}$ & $ 0.511^{+.060}_{-.060}$ &
$ 0.689^{+.150}_{-.050}$ & $ 0.687^{+.080}_{-.080}$ \\
~         &~ & ~ & ~ & \\
\hline
~ &\multicolumn{3}{c}{Pole position}& ~  \\
\hline
~         &~ & ~ & ~ & \\
II sheet
& $0.987^{+.005}_{-.005}$ & ~
& $0.989^{+.005}_{-.005}$ & ~ \\
~
&$-i(0.045^{+.005}_{-.005})$ & ~
&$-i(0.048^{+.010}_{-.010})$ & ~ \\
\hline
~         &~ & ~ & ~ & \\
III sheet
& $0.964^{+.015}_{-.015}$ & $1.558^{+.025}_{-.025}$
& $0.965^{+.015}_{-.015}$ & $1.571^{+.025}_{-.025}$ \\
~
&$-i(0.070^{+.010}_{-.010})$ & $-i(0.141^{+.015}_{-.015})$
&$-i(0.073^{+.010}_{-.010})$ & $-i(0.151^{+.015}_{-.015})$\\
\hline
~& \multicolumn{3}{c}{~} & ~\\
~& \multicolumn{3}{c}{$a_0$-resonances with K-matrix
background term} & ~ \\
\hline
~ & Solution I-1  & ~
& Solutions II-(1,2) & ~  \\
\hline
~ & $\alpha=1$ &$\alpha=2$  & $\alpha=1$ &$\alpha=2$   \\
\hline
~         &~ & ~ & ~ & \\
M   &
$0.944^{+.015}_{-.015}$ &$1.624^{+.100}_{-.030}$ &
$0.939^{+.015}_{-.015}$ &$1.640^{+.100}_{-.040}$ \\
~         &~ & ~ & ~ & \\
$g^{(\alpha)}$ &
$0.879^{+.100}_{-.100}$ &$ 0.702^*        $ &
$0.901^{+.100}_{-.100}$ &$ 0.435^*       $ \\
~         &~ & ~ & ~ & \\
$g_{4}^{(\alpha)}$&
$ 0.651^{+.100}_{-.080}$ & $ 0.519^{+.060}_{-.060}$ &
$ 0.653^{+.150}_{-.050}$ & $ 0.651^{+.080}_{-.080}$ \\
~         &~ & ~ & ~ & \\
~ & $f_{11}=0.529^{+100}_{-100}$ & $s_0=1.0^{+2.0}_{0.3}$
  & $f_{11}=0.731^{+100}_{-100}$ & $s_0=1.9^{+2.0}_{0.8}$ \\
~         &~ & ~ & ~ & \\
\hline
~ & \multicolumn{3}{c}{Pole position} & ~\\
\hline
~         &~ & ~ & ~ & \\
II sheet
& $0.990^{+.005}_{-.005}$ & ~
& $0.993^{+.005}_{-.005}$ & ~ \\
~
&$-i(0.039^{+.005}_{-.005})$ & ~
&$-i(0.042^{+.010}_{-.010})$ & ~ \\
\hline
~         &~ & ~ & ~ & \\
III sheet
& $0.965^{+.015}_{-.015}$ & $1.559^{+.025}_{-.025}$
& $0.965^{+.015}_{-.015}$ & $1.575^{+.025}_{-.025}$ \\
~
&$-i(0.063^{+.010}_{-.010})$ & $-i(0.145^{+.015}_{-.015})$
&$-i(0.068^{+.010}_{-.010})$ & $-i(0.153^{+.015}_{-.015})$\\
\hline
\end{tabular}
\end{table}
%---------------------------------------------------------------
%\newpage
\begin{table}
\caption\protect{Masses and coupling constants (in GeV) for
$f_2$ and $a_2$ resonances.}
\label{latetable}
\begin{tabular}{|l|cc|cc|}
\hline
~& \multicolumn{3}{c}{~} & ~\\
~ & \multicolumn{3}{c}{$f_2$-resonances} & ~  \\
\hline
~ & Solution I-1 & ~
& Solutions II-(1,2) & ~  \\
\hline
~ & $\alpha=1$ &$\alpha=2$  & $\alpha=1$ &$\alpha=2$   \\
\hline
~         &~ & ~ & ~ & \\
M   &
$1.236^{+.010}_{-.010}$ &$1.530^{+.010}_{-.010}$ &
$1.233^{+.010}_{-.005}$ &$1.529^{+.010}_{-.010}$ \\
~         &~ & ~ & ~ & \\
$g^{(\alpha)}$ &
$1.342^{+.100}_{-.100}$ &$ 1.342^{+.100}_{-.100}$ &
$1.038^{+.100}_{-.100}$ &$ 1.038^{+.100}_{-.100}$ \\
~         &~ & ~ & ~ & \\
$\Phi_\alpha$&
$-(8.4^{+2.0}_{-3.0})$ & $86.6^{+2.5}_{-4.5}$ &
$-(8.8^{+2.0}_{-3.0})$ & $86.2^{+2.5}_{-4.5}$ \\
~         &~ & ~ & ~ & \\
$g_{4\pi}^{(\alpha)}$&
$ 0.318^{+.020}_{-.020}$ & $ 0.448^{+.020}_{-.020}$ &
$ 0.318^{+.020}_{-.020}$ & $ 0.472^{+.020}_{-.020}$ \\
~         &~ & ~ & ~ & \\
\hline
~ & $a=\pi\pi$ & $a=\eta\eta$ & $a=\pi\pi$ & $a=\eta\eta$ \\
\hline
~         &~ & ~ & ~ & \\
$f_{1a} $ & $0.742^{+.050}_{-.250}$ &$-(1.01^{+.050}_{-.500})$
& $0.287^{+.070}_{-.070}$ &$-0.143^{+.100}_{-.100}$\\
~         &~ & ~ & ~ & \\
$r_{a} $ & $1.997^{+.150}_{-.150}$ &$1.077^{+.050}_{-.500}$
& $2.474^{+.150}_{-.150}$ &$1.295^{+.150}_{-.150}$\\
~         &~ & ~ & ~ & \\
~& $f_{13}=0.684\pm 0.100$ & ~&
$f_{13}=0.578\pm 0.100$& ~ \\
~& \multicolumn{3}{c}{$f_{ba}=0,\;\;b=2,3\;\; s_0=5.0$} & ~ \\
\hline
~         &~ & ~ & ~ & \\
Pole
& $1.262^{+.005}_{-.005}$ & $1.514^{+.010}_{-.006}$
& $1.261^{+.005}_{-.005}$ & $1.522^{+.005}_{-.010}$ \\
position
&$-i(0.092^{+.005}_{-.005})$ & $-i(0.066^{+.008}_{-.005})$
&$-i(0.089^{+.005}_{-.005})$ & $-i(0.076^{+.005}_{-.007})$\\
~         &~ & ~ & ~ & \\
\hline
\hline
~& \multicolumn{3}{c}{~} & ~\\
~ & \multicolumn{3}{c}{$a_2$-resonances} & ~  \\
\hline
~ & Solution I-1 & ~
& Solutions II-(1,2) & ~  \\
\hline
~ & $\alpha=1$ &$\alpha=2$  & $\alpha=1$ &$\alpha=2$   \\
\hline
~         &~ & ~ & ~ & \\
M   &
$1.316^{+.005}_{-.005}$ &$1.645^{+.050}_{-.050}$ &
$1.312^{+.005}_{-.005}$ &$1.695^{+.050}_{-.080}$ \\
~         &~ & ~ & ~ & \\
$g^{(\alpha)}$ &
$1.080^{+.100}_{-.100}$ &$ 0.270^{+.100}_{-.100}$ &
$1.300^{+.100}_{-.100}$ &$ 0.325^{+.100}_{-.100}$ \\
~         &~ & ~ & ~ & \\
$g_{4}^{(\alpha)}$&
$ 0.381^{+.050}_{-.050}$ & $ 0.597^{+.050}_{-.050}$ &
$ 0.426^{+.050}_{-.050}$ & $ 0.617^{+.050}_{-.050}$ \\
~         &~ & ~ & ~ & \\
~ & $r_1=1.845^{+.150}_{-.150}$& ~
& $r_1=2.406^{+.150}_{-.150}$ & ~ \\
\hline
~         &~ & ~ & ~ & \\
Pole
& $1.309^{+.005}_{-.005}$ & $1.615^{+.030}_{-.030}$
& $1.308^{+.005}_{-.005}$ & $1.667^{+.030}_{-.030}$ \\
position
&$-i(0.058^{+.005}_{-.005})$ & $-i(0.121^{+.015}_{-.015})$
&$-i(0.059^{+.005}_{-.005})$ & $-i(0.123^{+.015}_{-.015})$\\
~         &~ & ~ & ~ & \\
\hline
\end{tabular}
\end{table}

\end{document}